%% file: main.tex
\documentclass{IEEEtran}
\usepackage{cite}
\usepackage{amsmath,amssymb,amsfonts}
\usepackage{graphicx}
\usepackage{textcomp}
\usepackage{array}
\usepackage{threeparttable}
\usepackage{float}
\usepackage{lipsum}
\usepackage{lettrine} 
\usepackage{siunitx}
\usepackage{tabularx}
\usepackage{parskip}
\usepackage[caption=false]{subfig}
\usepackage{svg}
\usepackage{makecell}
\usepackage{multirow}
\usepackage{array}
\usepackage{booktabs}
\usepackage{acronym}
\usepackage{booktabs}
\usepackage{amsmath,amssymb}
\usepackage{mathtools} 
\usepackage{adjustbox}
\usepackage{cuted}

\allowdisplaybreaks        
\renewcommand{\textcolor}[2]{#2}

\usepackage[ruled,vlined,linesnumbered]{algorithm2e}
\DontPrintSemicolon
\SetAlgoNoEnd
\SetKw{KwAnd}{and}
\SetKw{KwOr}{or}
\SetKwRepeat{Repeat}{repeat}{until}

\newacro{EM}{electromagnetic}
\newacro{MoM}{method-of-moments}
\newacro{EFIE}{electric field integral equations}
\newacro{MPI}{message passing interface}
\newacro{GPU}{graphic processing unit}
\newacro{RWG}{Rao-Wilton-Gilson}
\newacro{BF}{basis function}
\newacro{FoM}{figure-or-merit}
\newacro{DBS}{direct binary search}
\newacro{RMS}{root mean square}
\newacro{RMSE}{root mean square error}
\newacro{DTW}{dynamic time warping}
\newacro{GA}{genetic algorithm}
\newacro{BEM}{boundary element method}
\newacro{PCB}{printed circuit board}
\newacro{UNII}{Unlicensed National Information Infrastructure}

\newcolumntype{Y}{>{\centering\arraybackslash}X}         
\newcolumntype{C}[1]{>{\centering\arraybackslash}p{#1}}  

\def\BibTeX{{\rm B\kern-.05em{\sc i\kern-.025emb}\kern-.08emT\kern-.1667em\lower.7ex\hbox{E}\kern-.125emX}}

\begin{document}
\title{Inverse Design of Multi-Layered Manufacturable Pixelated Diplexers Through Optimized Geometrical Configuration and Meshing Strategy in MoM}
\author{Woojun~Lee,~Jungmin~Lee,~Ji Wu~Hong~and~Jeffrey~Sean~Walling
\thanks{(Corresponding author: Woojun Lee.)}
\thanks{Woojun Lee, Jungmin Lee, Ji Wu Hong, and Jeffrey Walling are with the Bradley Department of Electrical and Computer Engineering, Virginia Polytechnic Institute and State University, Blacksburg, VA 24061 USA. (e-mail:woojun@vt.edu; jmlee22@vt.edu; hjiwu18@vt.edu; jswalling@vt.edu).}
}
\maketitle

\begin{abstract}
This paper presents a generalized white-box inverse design framework for multi-layered, multi-port pixelated diplexers based on a novel three-phase design flow including a selection of geometrical configurations including port placements, physical footprints, and microstrip/stripline configurations, a pre-computed \ac{MoM} formulation and a customized depth-increasing all-way tree search algorithm. The framework enables fast and rigorous optimization of complex three-dimensional electromagnetic (EM) structures by leveraging GPU-accelerated matrix reconstruction and selective basis-function assembly. The proposed search algorithm, extended from the \ac{DBS} method, dynamically explores multi-state pixel maps to minimize a composite figure-of-merit that captures insertion loss, isolation, and return loss across two passbands. The method is demonstrated through the design of several multi-layered pixelated diplexers. The resulting devices achieve as low insertion loss as 0.95--2.37~dB, high peak inter-channel rejection up to 78~dB, moderate worst inter-channel rejection as 30~dB and compact physical size as small as 0.16~$\times$~0.16~$\lambda_0$. Comprehensive analyses are conducted to evaluate the effects of substrate stack-up, finite grounds, and fabrication tolerances. This work also identifies mesh configurations for moderately accurate yet fast \ac{MoM} simulation. Fabricated prototypes based on standard PCB processes exhibit strong agreement between measurement and simulation, validating the manufacturability of the proposed approach. This work marks the first demonstration of multi-layered, multi-port pixelated diplexers.
\end{abstract}


\input{intro}

\input{sec2}

\input{sec_search}
\input{sec3}
\input{sec4}
\input{sec5}
\input{sec6}
\input{conclusion}

\bibliography{IEEEbibtex_v1}
\bibliographystyle{IEEEtran}

\begin{IEEEbiography}[{\includegraphics[width=1in,height=1.25in,clip,keepaspectratio]{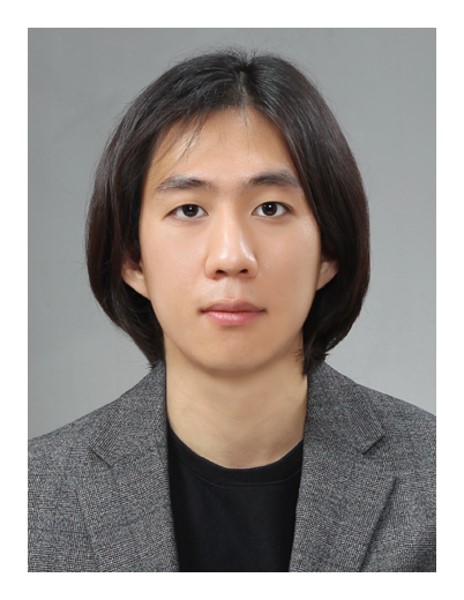}}]{Woojun Lee (Graduate Student Member, IEEE)}
received the B.S. and M.S. degrees in electrical and computer engineering from Seoul National University, Korea, in 2021 and 2023, respectively, and is currently pursuing the Ph.D. degree in electrical and computer engineering at Virginia Tech, Blacksburg, VA, USA. He is also currently working as a MCM/BAW design intern at Skyworks Solutions Inc., San Jose, CA, USA. \\ 
\hspace*{1em}His current research interests include inverse design in electromagnetics, active integrated antennas, and RF/microwave integrated circuits. \\
\hspace*{1em}He was a Qualcomm Innovation Fellowship North America 2024 Finalist and received the 2nd Prize in the IEEE MTT-S International Microwave Symposium (IMS) Student Design Competition for a PCB-based filter design.
\end{IEEEbiography}

\end{document}

%% file: intro.tex
\section{Introduction}
\label{sec:introduction}
\par \lettrine{O}{ptimization} methods have become essential for designing distributed \ac{EM} structures where closed-form mathematical expression is limited to design such structures. Traditional heuristic approaches—such as genetic algorithms and particle swarm optimization \cite{psogenetic}—have been widely used to optimize \ac{EM} structures by systematically exploring geometric parameters. However, these techniques heavily rely on computationally expensive commercial \ac{EM} simulations, limiting their scalability and efficiency. On the other hand, the optics community has been applying inverse design to silicon photonics \cite{deep_learning_optics, objective_first}, encoding device geometries into programmable surfaces and optimizing them with the state-of-the art optimization techniques such as neural networks or convex optimization. 

Following this direction, the microwave community has started to explored more sophisticated optimization strategies for "pixelated" microwave components, including antennas \cite{gamom, sideris2}, filters \cite{pixelfilterREF1, pixelfilterREF2, pixelfilterREF3, jungmin, reviewrREF1, reviewrREF2}, and matching transformers/networks in integrated circuits \cite{utaustin, usc, princetonIMS, huawang}. Two primary approaches have emerged: neural-network-based emulation and computational \ac{EM} acceleration. Machine learning-driven EM emulators are proposed to replace time-consuming \ac{EM} simulations in \cite{princetonIMS}. In \cite{huawang, utaustin, usc}, on-chip transformers were optimized using neural networks. \textcolor{blue}{In \cite{reviewrREF1}, a convolutional neural network is combined with a deep reinforcement learning agent to optimize substrate-integrated waveguide filters in a hybrid binary–continuous design space.}  On the other hand, pre-computation-based solvers have enabled rapid evaluation of pixelated \ac{EM} devices. In \cite{gamom}, the pre-computation method in a method-of-moments (MoM) formulation is coupled with \ac{GA} to optimize pixelated antennas. Recently, pre-computation using custom finite-difference and \ac{BEM} formulations to accelerate the full-wave simulation has been applied to mmWave antennas and beam-switching antennas in \cite{sideris2}. In \cite{jungmin}, a simple, yet very efficient \ac{DBS} method is used for efficient exploration of discrete pixel states to synthesize a dual-band filters on \ac{PCB}, which are widely used in \ac{EM} inverse design.

\begin{figure*}[t!]
    \centering
    \includegraphics[width=\textwidth]{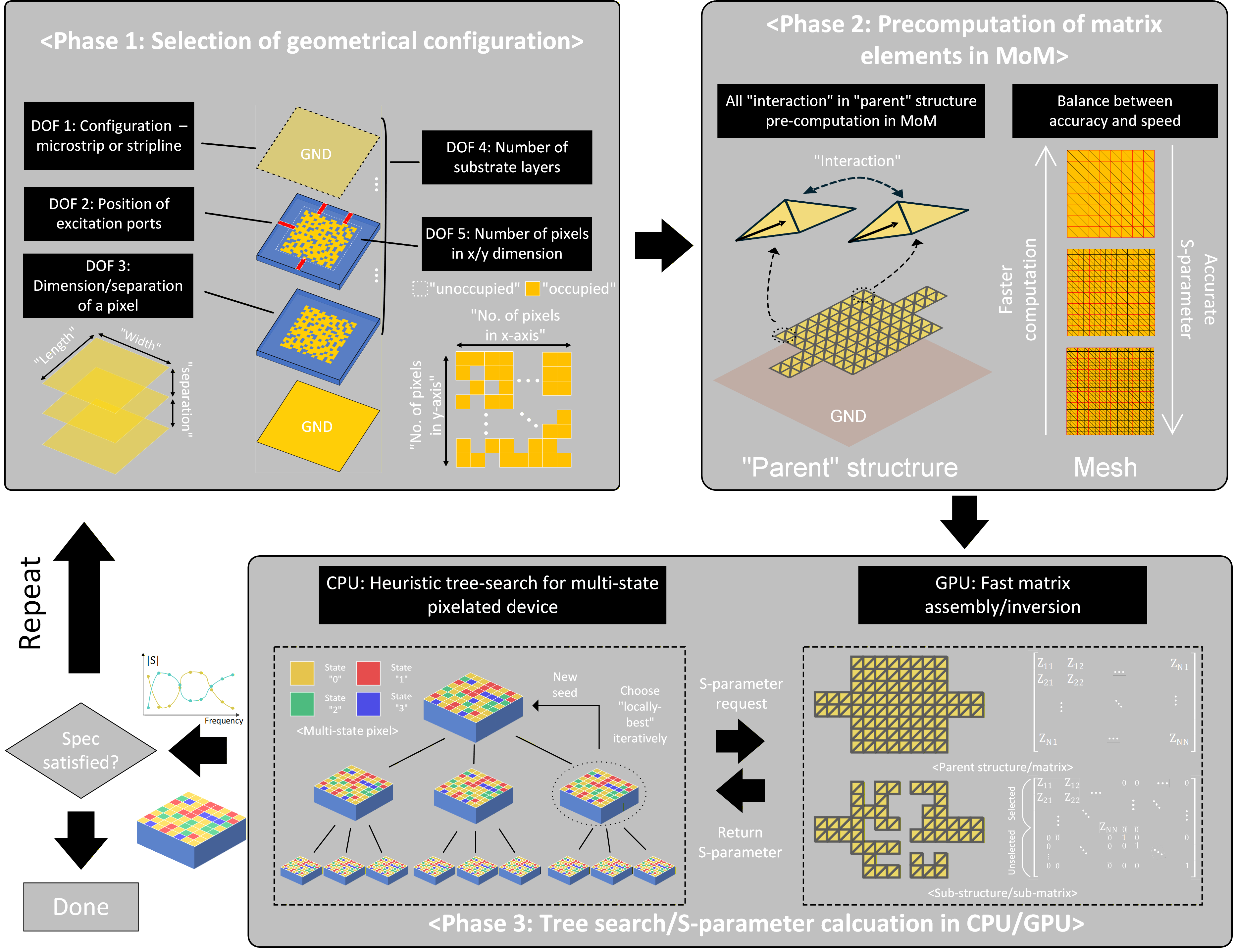}
    \caption{The prposed three-phase inverse-design workflow for multilayer, multi-port pixelated electromagnetic devices. Phase 1 defines the geometric configuration, including a signal/ground configuration, port placement, substrate stack-up, and pixel resolution. Phase 2 performs \ac{MoM}-based pre-computation of interaction matrices per frequency. Phase 3 applies an iterative tree-search optimization to update pixel states while matrix inversion is offloaded to \ac{GPU}s.}
    \label{fig::FIG1} 
\end{figure*}

Despite these advances, most prior work remains focused on single-layer structures, typically limited to two-port devices such as filters or matching networks, without exploiting the compactness and design freedom offered by multilayer implementations \cite{mutlilayered_1}. Realizing the full capability of inverse design requires synthesizing multi-port, multilayered structures that are difficult to conceive manually. Although some efforts have applied inverse design to multilayer on-chip transformers \cite{utaustin, usc}, those designs are constrained to two port structure yet. In addition, while recent studies have demonstrated pixel-based filters \cite{pixelfilterREF1, pixelfilterREF2, pixelfilterREF3}, no prior work has reported pixelated diplexers, leaving multi-port inverse design largely unexplored.

This paper addresses this gap by introducing a fast inverse design framework based on a tree search algorithm coupled with a \ac{MoM} solver. This approach tackles three central challenges inherent in multilayered EM device design. First, the computational demands increase dramatically with the addition of layers. Memory requirements grow quadratically due to the expanding matrix size from the increased number of \ac{BF}. Moreover, computational time for matrix inversion increases cubically with each new layer. Second, achieving an optimal balance between computational speed and accuracy hinges on judicious selection of \ac{BF}. An overly fine resolution leads to excessive computational time, while a coarse resolution compromises the accuracy of the results. Finally, search complexity problem is tackled with a customized depth-increasing all-way navigating tree search with gradually increasing depth. 

Furthermore, none of the existing inverse design studies have thoroughly investigated the impact of manufacturing tolerances that occur during the etching process or due to variations in substrate thickness. In particular, most works on pixelated configurations do not specify whether diagonally adjacent pixels are considered electrically connected, which can significantly affect the resulting performance. This work addresses these practical limitations by analyzing both under-etch and over-etch conditions and quantifying the resulting \ac{RMS} errors. In addition, a detailed parametric study is conducted to evaluate the influence of slight variations in substrate thickness and conductor heights on the frequency responses of multilayered diplexers.

The contributions of this work compared to previous studies on inverse design of pixelated surfaces such as \cite{huawang, jungmin, princetonIMS, sideris_ims, pixelfilterREF1, pixelfilterREF2, pixelfilterREF3, usc} are as follows. 1) Generalized multi-port, multi-layered extensions of the \ac{MoM} matrix pre-computation with \ac{MPI}, \ac{GPU} accelerated work flow. To the author's knowledge, this is the first time that two-layered, triple-layered multi-port inverse design is proposed in a microwave community. 2) Expansion of direct-binary search algorithm proposed in \cite{jungmin} to a depth-increasing all-way navigating tree-search, that allows the optimization of multi-layered much more robustly. 3) Comprehensive field analyses are conducted on the designed pixelated diplexer to reveal the underlying electromagnetic behaviors and coupling mechanisms within the structure. 4) Study on a meshing strategy including the number of triangles and orientation distribution in \ac{MoM} to find the balance between speed and accuracy. 5) Study on manufacturing tolerances in the designed pixelated diplexer, including different substrate stack-up, under-/over-etching, and conductor thickness deviations. This is the first time that such tolerance studies are carried out for inversely designed pixelated devices.

This paper is organized as follows. Section~\ref{sec:design_method} introduces the proposed multi-phase inverse design framework, including the formulation of the pre-computed \ac{MoM} database, basis-function classification, and an extension into multi-layered multi-port design. Section~\ref{sec:optim_phase} proposes a depth-increasing all-way tree-search algorithm with a comparison with \ac{GA} and a definition of \ac{FoM} for a diplexer design with a target specification. Section~\ref{sec::sec3} presents the optimized pixelated diplexer results under various configurations, including microstrip and stripline implementations, together with detailed field analyses explaining their electromagnetic behavior. Section~\ref{sec::sec4} discusses the meshing strategy used in the MoM solver, identifying the optimal mesh density and orientation through systematic \ac{RMSE} evaluations. Section~\ref{section::Manufacture_error} investigates the impact of practical manufacturing tolerances such as under-etching, stack-up variation, conductor dimensionality, and finite ground size on the device performance. Section~\ref{sec::measurement} reports the fabrication and measurement results of the proposed diplexers, highlighting their strong consistency with simulations and comparing their performance against state-of-the-art designs, covering both diplexer performance metrics and the computational cost associated with the optimization process. Finally, Section~\ref{sec:conclusion} concludes the work with future research directions.

%% file: sec2.tex
\section{Three-phase Flow and Pre-Computation Phase}
\label{sec:design_method}

This section discusses the proposed three-phase design flow, ranging from choosing the geometrical configuration to the customized optimization engine, which is based on a pre-computation based accelerated electromagnetic simulation and the depth-increasing all-way tree search algorithm.

\subsection{Three-phase Design Flow}
\label{subsection::design_flow}
Fig. \ref{fig::FIG1} presents the three-phase design flow used to achieve the specified electromagnetic response in multi-state, multilayer pixelated devices. In phase 1, the basic geometrical configuration is determined, including the transmission-line type (microstrip or stripline), the position of excitation ports, the number of substrate layers, and the pixel resolution in the lateral dimensions. The size and arrangement of pixels define the degrees of freedom available during optimization and determine how finely the electromagnetic field distribution can be shaped within the structure. This step also sets the boundary conditions and the overall stack-up before numerical analysis begins.

In phase 2, numerical interactions among all \ac{BF}s are computed using \ac{MoM} for a reference or “parent” structure, in which all the conducting pixels are occupied. The resulting interaction matrix contains all the needed matrix entry to simulate the "sub"-structure, where a part of the conducting pixels are occupied or drawn. A balance between mesh density and computation time is maintained to achieve both acceptable accuracy and manageable resource usage. The precomputed matrices are later called upon for fast GPU-based matrix assembly and inversion, significantly accelerating the repeated analyses required in the design cycle.

Phase 3 involves the iterative optimization process, where the pixel states are updated through the depth-increasing all-way navigating tree-search algorithm extended from \ac{DBS} method. For each candidate configuration, the corresponding $S$-parameters are computed, and the results are compared with the target specification. 

If the desired response is not yet achieved, a new configuration is generated and evaluated. This loop continues until the specifications are satisfied, resulting in a final geometry that meets the performance targets with high computational efficiency. The following subsections will focus on the pre-computation phases for multi-layered multi-port electromagnetic devices.

\subsection{MoM Background}
\label{subsection::background}
This subsection briefly discusses the \ac{MoM} in \ac{EM}. One approach to deriving the frequency response of EM devices involves solving the electric field integral equation (EFIE). This method utilizes an equivalent representation of the structure based on the distribution of current density on conducting surfaces. The \ac{EFIE} is expressed as:
\begin{equation}
\hat{n} \times \vec{E}_{\text{inc}} - \hat{n} \times \iint_S \overline{\overline{G}}(\vec{r}, \vec{r}') \cdot \vec{J}(\vec{r}') \, dS' = 0
\end{equation}
where $\vec{E}_{\text{inc}}$ is the incident electrical field, $\vec{J}(\vec{r}')$ is the surface current density, $\hat{n}$ is the normal vector to the conducting surface, and $\overline{\overline{G}}$ is the dyadic Green's function, which depends on the dielectric property and configuration of the surfaces. 

\ac{MoM} is a well-established technique for solving the EFIE \cite{mom}. By discretizing the conducting surface into smaller meshes and approximating the surface current density using a set of basis functions (BFs), the EFIE is transformed into a matrix equation:
\begin{equation}
    \mathbf{Z}\mathbf{I} = \mathbf{V}
    \end{equation}
where $\mathbf{Z}$ is an interaction matrix that specifies the relationship between the BF representing the surface current density of the conducting surfaces, $\mathbf{V}$ is an excitation vector representing the source, and $\mathbf{I}$ is a vector containing the coefficient of each BF to be solved. In this paper, \ac{RWG} \ac{BF} is employed \cite{rwg}, where each edge of a triangular mesh defines a \ac{BF}. 


\subsection{Pre-computation with Basis Function Classification}

\begin{figure}[t]
    \centering
    \includegraphics[width=1\columnwidth]{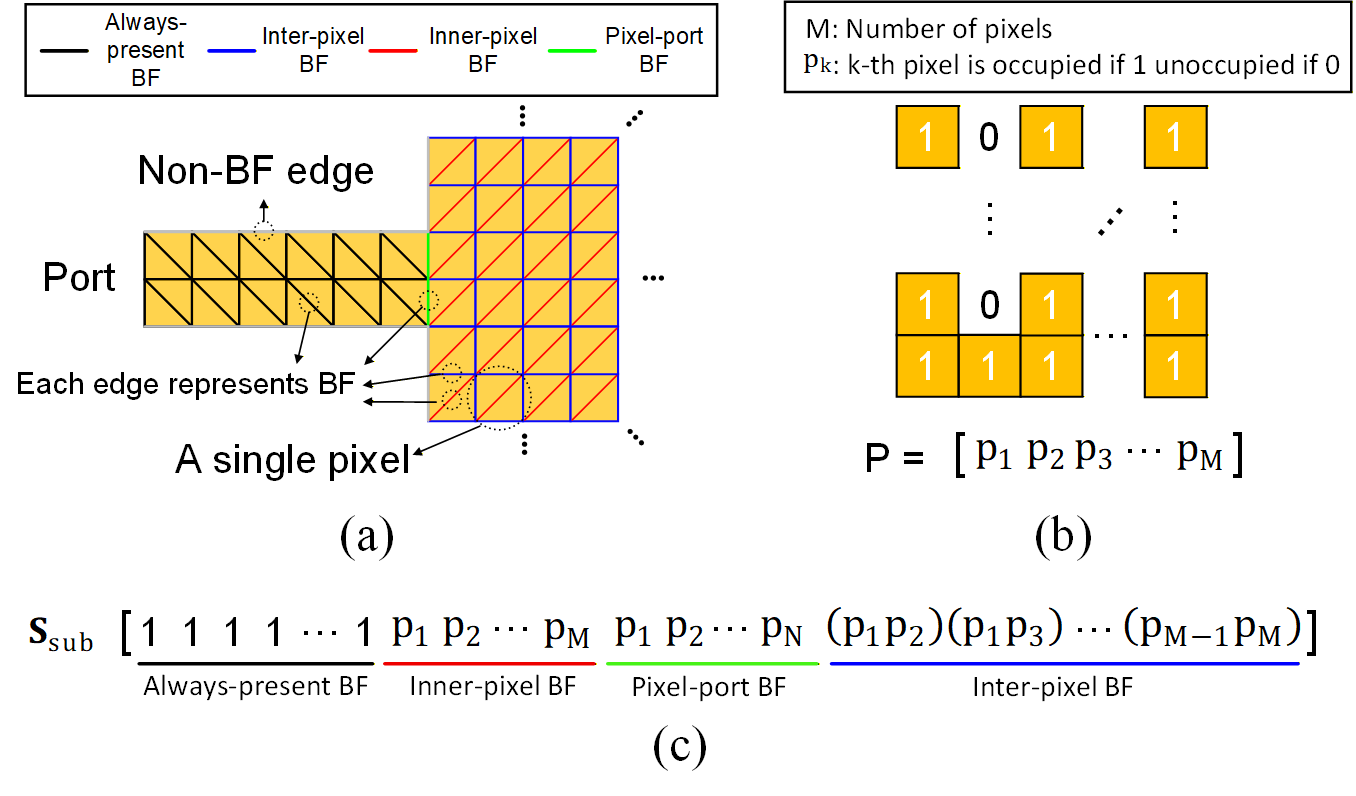}
    \caption{(a) Classification of basis functions into "inter-pixel," "inner-pixel," and "always-present" categories. (b) Construction of a vector $P$ that represents the occupation of the conducting pixels. (c) Construction of a selection vector $S$ to reconstruct a sub-interaction matrix that represents the structure.}
\label{fig::rwg_classifcation}
\end{figure}

The proposed method extends the previous approach of pre-computing the matrices in \ac{MoM} in \cite{gamom} to accommodate multi-layer, multi-port metal–dielectric composite structures. Similar approach is reported in \cite{gamom} to construct the impedance matrix of a sub-structure by selectively removing rows and columns from the original "parent" matrix for a pixelated metallic antenna. It is the first time to be employing this method to optimize multi-layered multi-port electromagnetic scattering surfaces to deliver the desired multi-port S-parameter matrix, not optimizing a single-port impedance.

The reconstruction procedure of the interaction matrix for the substructure is described in detail. First, each \ac{BF} is pre-classified, as illustrated in Fig.~\ref{fig::rwg_classifcation}(a). Four main categories are defined: “always-present,” “inner-pixel,” “inter-pixel,” and “pixel-port” \ac{BF}s. The always-present \ac{BF}s correspond to \ac{BF}s that remain active regardless of pixel states—typically along the feed or launching section. Inner-pixel \ac{BF}s are associated with edges located entirely within a single pixel and are activated only when that pixel is in the “on” state. Inter-pixel \ac{BF}s lie between adjacent pixels and are included in the impedance matrix only when both neighboring pixels are active. In addition, pixel-port BFs are introduced to accurately capture current flow between the launching traces and the adjacent active pixels.

Based on the classification described above, the square interaction matrix
$\mathbf{Z}_{\mathrm{sub}}$ corresponding to a substructure of the fully
occupied parent structure is reconstructed from the parent interaction
matrix $\mathbf{Z}_{\mathrm{parent}}$ using a selection vector
$\mathbf{S}_{\mathrm{sub}}$, as illustrated in Fig.~\ref{fig::rwg_classifcation}.
The support index set $\mathcal{I}$ is defined as
\begin{equation}
\mathcal{I} \triangleq \{\, i \in \{1,\dots,M\} \mid (\mathbf{S}_{\mathrm{sub}})_i = 1 \,\}.
\end{equation}
where $M$ denotes the dimension of $\mathbf{Z}_{\mathrm{parent}}$ and
$N = |\mathcal{I}|$. Let $\mathcal{I} = \{ i_1, i_2, \dots, i_N \}$ with
$i_1 < i_2 < \cdots < i_N$.
The selection matrix $\mathbf{S}_{\mathrm{mat}} \in \mathbb{R}^{M \times N}$
is defined element-wise as
\begin{equation}
\left( \mathbf{S}_{\mathrm{mat}} \right)_{m,n}
\triangleq
\delta_{m,i_n},
\qquad
m = 1,\dots,M,\;\;
n = 1,\dots,N,
\end{equation}
where $\delta_{m,i_n}$ denotes the Kronecker delta.
Finally, the reduced interaction matrix
$\mathbf{Z}_{\mathrm{sub}}$ is obtained as
\begin{equation}
\mathbf{Z}_{\mathrm{sub}}
=
\mathbf{S}_{\mathrm{mat}}^{\mathsf T}
\mathbf{Z}_{\mathrm{parent}}
\mathbf{S}_{\mathrm{mat}}.
\end{equation}
This formulation enables extract the rows and columns of
$\mathbf{Z}_{\mathrm{parent}}$ corresponding to the active pixels specified
by $\mathbf{S}_{\mathrm{sub}}$.

\subsection{Extension to Multi-layered Design}
\label{subsection::multi_layered}

\begin{figure}[t]
    \centering
    \includegraphics[width=1\columnwidth]{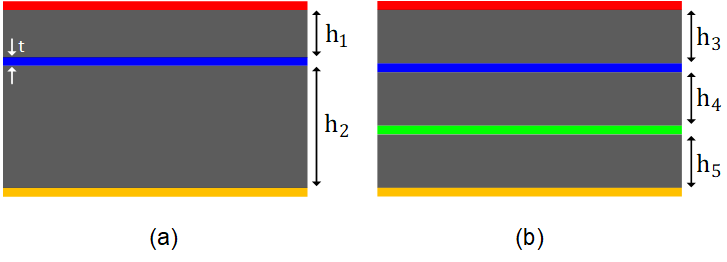}
    \caption{Substrate stack-up used for this work. (a) Non-uniform three-layer stack-up. (b) Uniform/non-uniform four-layer stack-up which can be used for both microstrip and stripline configurations.}
\label{fig::substrate_setup}
\end{figure}

\begin{figure}[t]
    \centering
    \includegraphics[width=1\columnwidth]{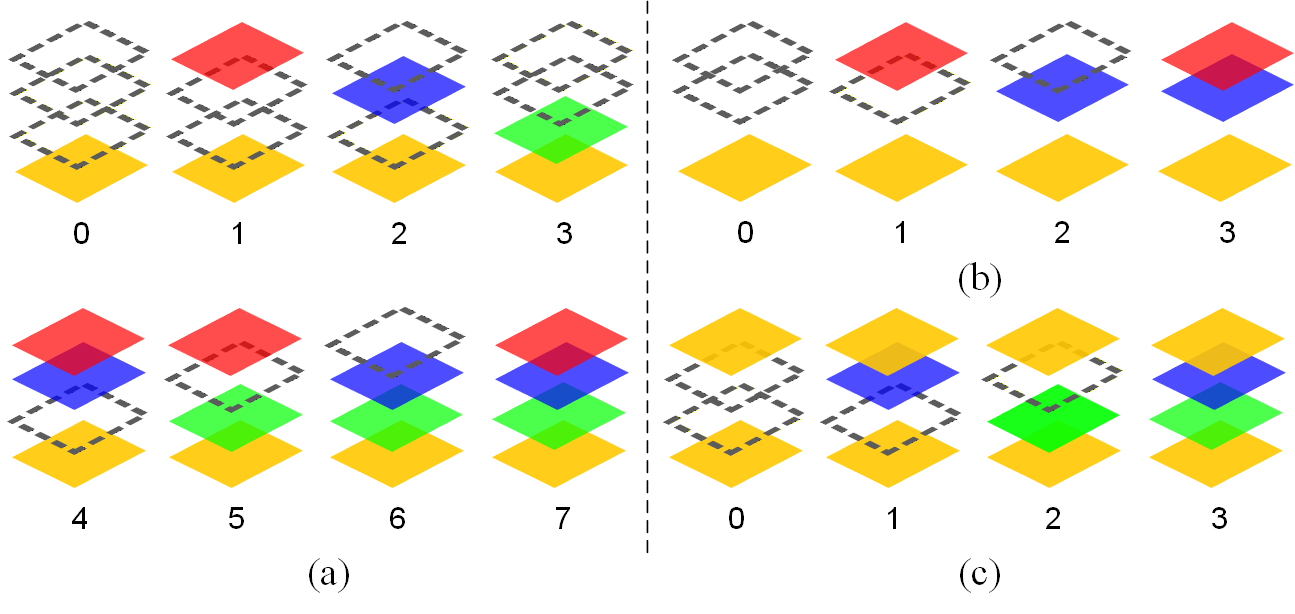}
    \caption{The proposed encoding of the pixel with corresponding configurations. (a) 8 States in microstrip. (b) 4 states in microstrip. (c) 4 states in stripline (Ground plane is marked as the different color from Fig.~\ref{fig::substrate_setup} (b)).} 
\label{fig::pixel_states}
\end{figure}

\textcolor{blue}{This subsection describes the definition of the design spaces associated with a multi-state pixelated design, based upon an addition of conduction layers to the conventional single-layered inverse design space. Such an extension will add up to richer design space by adding up vertical coupling between signal layers where the single-layered design relies only on co-planar coupling mechanism.} 

Fig.~\ref{fig::substrate_setup} illustrates two examples of substrate configurations used in the proposed design framework. In Fig.~\ref{fig::substrate_setup}(a), a two-layer structure represents a typical microstrip configuration, where the signal trace lies on the top layer and the lower metallic layer serves as the ground plane. In contrast, Fig.~\ref{fig::substrate_setup}(b) shows a three-layer example that can function either as a microstrip or as a stripline, depending on how the upper and lower conductors are assigned. In this configuration, the top red trace can be used as a signal layer or alternatively connected to ground. \textcolor{blue}{Fig.~\ref{fig::pixel_states} shows representative multi-state pixel configurations, where colors denote different conductive layers. Fig.~\ref{fig::pixel_states}(a) and (b) correspond to eight- and four-state cases, respectively, while Fig.~\ref{fig::pixel_states}(c) illustrates a grounded four-state configuration. These states define the discrete design space used during optimization.}

In \ac{MoM}'s perspective, mixed-potential integral formulation along with Sommerfeld-type integral in a stratified medium is employed \cite{MoM_multilayered}. The dielectric substrate in all configurations is modeled as a homogenized medium with a complex dielectric constant. It is noted that an infinite lateral dimension of the dielectric substrate is assumed both in microstrip and stripline configurations. The influence of finite ground size will be analyzed in the subsequent section. In this work, Altair FEKO was employed as the \ac{MoM} solver \cite{FEKO}.

\subsection{Extension to Multi-port Design}
\label{subsection::multi_port}
In this subsection, we explore how multi-port S-parameter is calculated in MoM solver. Each port node corresponds to one of the basis function discussed in the previous section. A general mathematical expression for each scattering parameter $S(i, j)$ is given as below \cite{mom_sparam}
\begin{equation}
   S(i, j) = \delta_{ij} - 2 Z_0 I_i / V_j 
\end{equation}
where $Z_0$ denotes the characteristic impedance of each port (assuming that the same characteristic impedance for all ports), $V_j$ denotes the voltage that is incident into the port j, and $I_i$ is the reflected current into the port $i$. $2$ in nominator is due to the voltage division between the port load and the transmission line assuming that both impedance are the same and minus sign is to take the direction of the current into account. Since \ac{RWG} basis functions are employed in this work, $I_i$ is further expressed as below 
\begin{equation}
    I_i = l_i * i_i
\end{equation}
where $l_i$ is a length of the common edge between two triangles and $i_i$ is the resulting coefficient of the basis function.

%% file: sec_search.tex
\section{Optimization Phase}
\label{sec:optim_phase}

\begin{algorithm}[t]
\caption{Depth-Increasing All-way Tree Navigation for Multi-State Pixelated Surface Optimization}

Initialize the root $\text{node}_{0,1}=\{\emptyset,\,P,\,F\}$\;
Set the initial navigating depth $D$ and define $M(D)$ as the maximum count allowed at each $D$ before increasing $D$\;
Set the total number of $F$ evaluations\;

\textbf{repeat:} depth-increasing all-way tree navigation\;
\Indp
  \textbf{repeat:} tree navigation to depth $D$\;
  \Indp
    From the root node, expand sequentially for $d=0,1,\dots,D{-}1$\;
    For each $\text{node}_{d,k}$, choose a pixel index $j\notin S_{d,k}$ and generate all local state changes for $j$\;
    Each change produces a child 
    $\text{node}_{d+1,\ell}=\{S_{d,k}\cup\{j\},\,P_{d+1,\ell},\,F_{d+1,\ell}\}$, 
    where $P_{d+1,\ell}$ differs from $P_{d,k}$ only at index $j$; evaluate $F_{d+1,\ell}$ for all children\;
    After all children at depth $D$ are evaluated, identify the locally best leaf
    $\{S_{\text{best}},\,P_{\text{best}},\,F_{\text{best}}\}$ with $\{P_{\text{best}},F_{\text{best}}\}=\arg\min_{\ell}F_{D,\ell}$\;
  \Indm
  \textbf{until} all nodes up to depth $D$ are explored\;

  If $F_{\text{best}}<F$ then update the current state $\{P,F\}\leftarrow\{P_{\text{best}},F_{\text{best}}\}$ and set the new root as $\{\emptyset,\,P,\,F\}$; otherwise increase a non-improvement counter\;
  If the counter $\ge M(D)$ then set $D=D+1$ and reset the counter\;
  If $D$ reaches the predefined maximum, initialize $P$ randomly, evaluate its $F$, reinitialize the root $\text{node}_{0,1}$ accordingly and reset $D$ to its initial value\;
\Indm
\textbf{until} the maximum number of F evaluations is reached\;
\end{algorithm}

\subsection{Depth-increasing All-way Navigating Tree Search}
\label{subsection::tree_search}

To ensure the robustness and convergence of the inverse design process, a customized version of the \ac{DBS} algorithm, referred to as an in-depth tree-search algorithm, is employed. As shown in the phase 3 in Fig.~\ref{fig::FIG1}, this method iteratively refines a multi-state pixel map to optimize a mono-objective \ac{FoM}, which reduces a complicated multi-objective problem to a mono-objective problem. The \ac{FoM} formulation, describing how effectively the design satisfies target specifications such as return loss, insertion loss, and isolation, will be detailed in the following subsection.

The algorithm starts with an initial pixel map. Before the search begins, one should define a tree node at depth $d$ and position $k$ as $\text{node}_{d,k}={S_{d,k},,P_{d,k},,F_{d,k}}$, where $S$ is the set of pixel indices chosen along the path, $P$ is the multi-state pixel map, and $F$ is its figure of merit. Each node represents a possible configuration in the search space. During each iteration, the algorithm selects an index and evaluates all possible directions for modification. Because each pixel can take multiple discrete states rather than binary ones, the search naturally expands into multiple branches. Each variation forms a new node, branching out from its parent node, and subsequent evaluations proceed independently for each branch. This process constitutes one full iteration of the algorithm.

If the \ac{FoM} shows no improvement after a predefined number of iterations, the search depth is increased to allow exploration of more complex pixel combinations. The maximum allowable depth and iteration threshold are pre-defined to prevent excessive computational time during a single run. Once the algorithm reaches saturation at a given depth, it restarts from a new random seed. This adaptive, hierarchical strategy enables the algorithm to explore the design space thoroughly, ultimately yielding stable convergence toward an optimized pixelated structure.

\subsection{\ac{FoM} with Target Specifications}
\label{subsec::target_specs}

The diplexers are designed to operate with one channel covering the \ac{UNII}-1–4 band (5.15–5.925~GHz) and the other covering the \ac{UNII}-6–8 band (6.425–7.125~GHz), within a compact footprint of $0.16\lambda_0 \times 0.16\lambda_0$. The primary design objective is insertion loss and return loss. Another objective includes, inter-channel rejection, defined as $S_{mn}$ in each channel, where port $m$ is the common (input) port and port $n$ is the unused port for a given channel. The target inter-channel attenuation is set as below $-20$~dB across each passband, which was challenging to achieve within the compact area while maintaining return losses below $-10$~dB at all three ports. The isolation between output ports is also targeted to remain below $-20$~dB over the operating bands.

Therefore, the \ac{FoM} employed to guide the optimization process is formulated as:

\newcommand{\Ss}{\{s_3,s_4\}}
\newcommand{\So}{\{m_1,m_2\}}
\newcommand{\Sabs}[2]{\lvert S_{#1#2}(f)\rvert}

\begin{equation}
\begin{aligned}
\mathrm{FoM} &=
  \sum_{f\in f_{p1}} (\frac{1}{rw_{p1}(f)}+\frac{1}{rw_{m1}(f)}+\frac{1}{rw_{s2}(f)}) \\
&\quad+\sum_{f\in f_{p2}} (\frac{1}{rw_{p2}(f)}+\frac{1}{rw_{m2}(f)}+\frac{1}{rw_{s1}(f)}) \\
&\quad+\sum_{f\in f_{i}}  \frac{1}{rw_{oi}(f)} + \sum_{f\in f_{s}} \sum_{t\in \Ss} \frac{1}{rw_{t}(f)} 
\end{aligned}
\label{eq:fom}
\end{equation}

\newcommand{\Sdb}[2]{S_{#1#2,\mathrm{dB}}(f)}


{\footnotesize
\begin{equation}
\label{eq:rw_defs}
\begin{aligned}
rw_{p1}(f) &= 10^{\tfrac{\Sdb{2}{1}+1.0}{10}}, &
rw_{p2}(f) &= 10^{\tfrac{\Sdb{3}{1}+1.0}{10}}, \\
rw_{s1}(f) &= 10^{\tfrac{-\Sdb{2}{1}-35}{15}}, &
rw_{s2}(f) &= 10^{\tfrac{-\Sdb{3}{1}-35}{15}}, \\
rw_{s3}(f) &= 10^{\tfrac{-\Sdb{2}{1}-20}{18}}, &
rw_{s4}(f) &= 10^{\tfrac{-\Sdb{3}{1}-20}{18}}, \\
rw_{oi}(f) &= 10^{\tfrac{-\Sdb{3}{2}-20}{15}}, &
rw_{m1}(f) &= 10^{\tfrac{-15-\Sdb{1}{1}}{10}}, \\
rw_{m2}(f) &= 10^{\tfrac{-15-\Sdb{1}{1}}{10}}.
\end{aligned}
\end{equation}
}

The \ac{FoM} in Eq.~(\ref{eq:rw_defs}) guides the optimization, where $f_{p1}$ represents the frequencies in the channel 1, $f_{p2}$ represents the frequencies in the channel 2, $f_{i}$ represents the intermediate frequencies between the channel 1 and channel 2, and $f_s$ denotes the other undefined frequencies from 3 GHz to 9 GHz. The terms $rw_{p1}(f)$ and $rw_{p2}(f)$ enhance passband transmission, $rw_{s1}(f)$--$rw_{s4}(f)$ enforce stopband rejection, $rw_{oi}(f)$ denotes isolation in the intermediate band, and $rw_{m1}(f)$ and $rw_{m2}(f)$ enhance return loss.

Each reward term is formulated as an exponential barrier, which promotes a fast identification of the desired frequency response from the random seed. The exponent coefficients in each rewarding function were empirically tuned with an emphasis in the order of insertion loss, return loss, inter-channel rejection, and out-of-band rejections. Each exponent may be tuned as the target specifications are altered. \textcolor{blue}{Impact of weight adjustment will be discussed in Section \ref{subsec::fom_weight}}.

\begin{figure}[t]
    \centering
    \includegraphics[width=1\columnwidth]{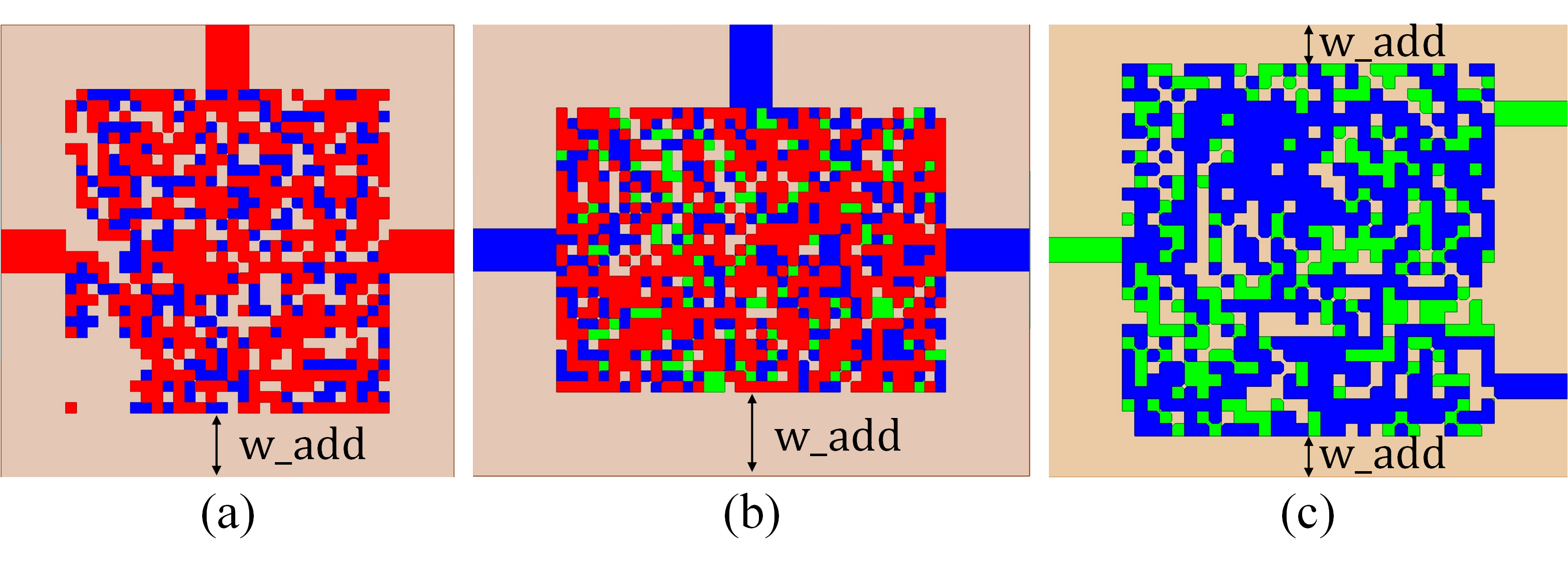}
    \caption{Exemplar diplexers with a proposed geometrical configuration and port positions on a finite ground. (a) 4 states in microstrip. (b) 8 states in microstrip. (c) 4 states in stripline.}
\label{fig::tolerance_finite_ground}
\end{figure}

\subsection{Comparison with GA}
\label{subsection::speed_up}
\textcolor{blue}{In this subsection, an experiment with the proposed tree search algorithm and \ac{GA} to compare average number of function evaluations and the final performance of the designed diplexers. Example pixel geometries are shown in Fig.~\ref{fig::tolerance_finite_ground}. The development of the geometrical configurations and parameters will be detailed in the following section. The 8-state pixel configuration in Fig. \ref{fig::tolerance_finite_ground}(b) is chosen in this subsection since it provides the most harsh environment of so many parasitic resonances. }

Hyper-parameters of each algorithm is described. The maximum depth of the tree search was set to 2, as each node requires an extensive 7-way branching over the remaining pixel states. The count that tolerates non-improvement was set to 10 consecutive iterations at depth 1 and reduced to 3 iterations at depth 2. The implementation of \ac{GA} to optimize 8-state pixelated devices is described. Each individual in the population represents a 8-state pixel map, which is mapped to a three-layer discrete layout and evaluated. Tournament selection with elitism is adopted, where the best individuals are preserved while parents are selected from small randomly sampled subsets of the population \cite{GA_goldberg}. Offsprings are generated through uniform crossover followed by probabilistic per-pixel mutation. The \ac{GA} in this work operates with a fixed population size of 1024 and does not have any termination conditions. 

\begin{figure}[t!]
    \centering
    \includegraphics[width=1\columnwidth]{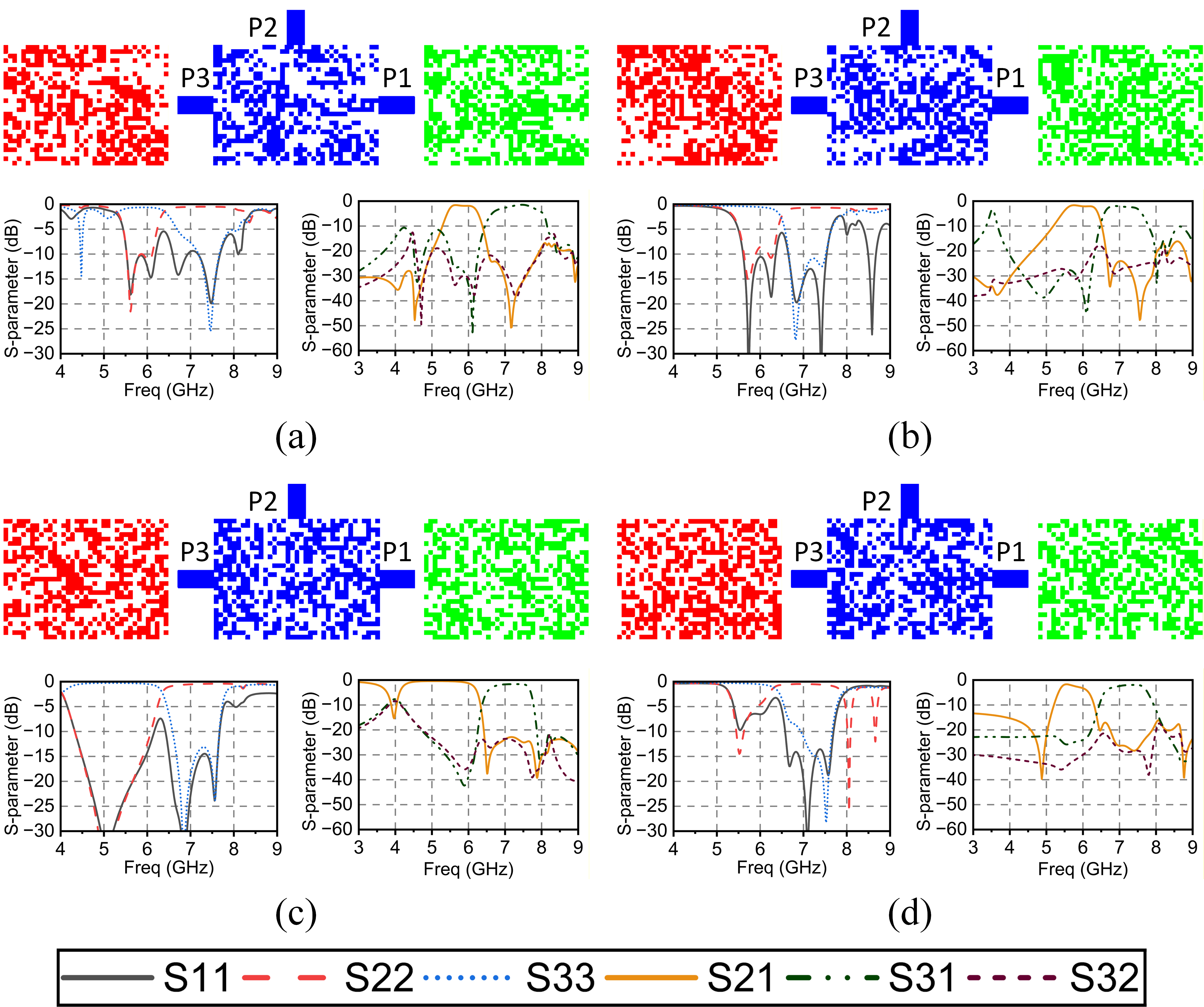}
    \caption{\textcolor{blue}{Exemplar diplexers optimized with two different optimizers on the 8-state microstrip configuration in Table \ref{tab:configuration_summary}. (a)-(b) Diplexers optimized with the proposed tree search algorithm. (c)-(d) Diplexers optimized with \ac{GA}.}}
\label{fig::pm_from_tree_and_GA}
\end{figure}

\begin{figure}[t!]
    \centering
    \includegraphics[width=1\columnwidth]{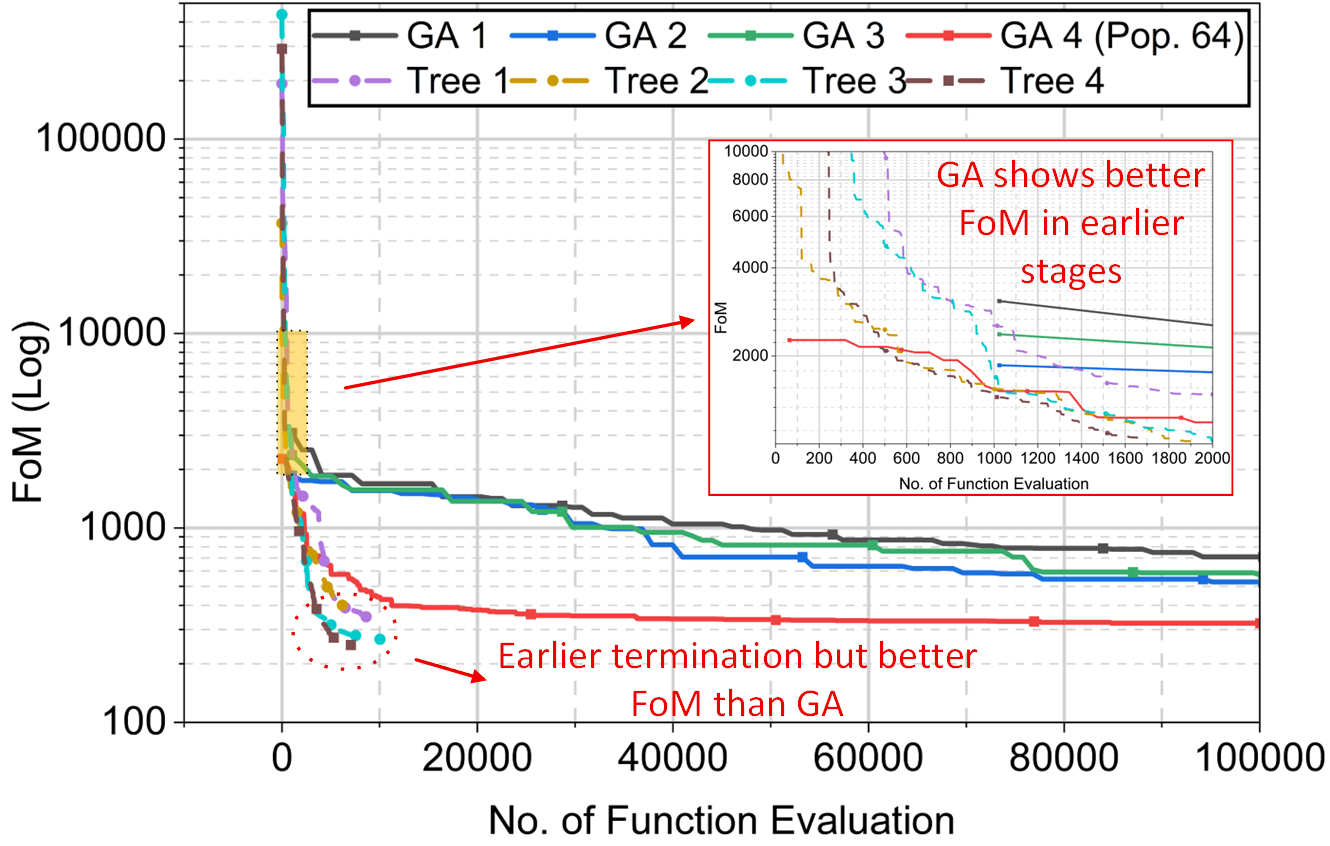}
    \caption{\textcolor{blue}{\ac{FoM} versus function evaluations for the \ac{GA} and the proposed tree-search algorithm. Tree~2–3 correspond to Fig.~\ref{fig::pm_from_tree_and_GA}(a)–(b). GA~1 (population size 1024) corresponds to Fig.~\ref{fig::pm_from_tree_and_GA}(c), and GA~4 (population size 64) corresponds to Fig.~\ref{fig::pm_from_tree_and_GA}(d). Results of Tree~1,4 and GA~2–3 (population size 1024) are not shown.}}
\label{fig::tree_vs_ga}
\end{figure}

Diplexers optimized using the proposed tree-search algorithm are shown in Fig.~\ref{fig::pm_from_tree_and_GA}(a)-(b), while those obtained using \ac{GA} are presented in Fig.~\ref{fig::pm_from_tree_and_GA}(c)-(d). In both optimizations, \ac{FoM} described in the previous sections are used. The performance \textcolor{blue}{of the optimizer} comparison is shown in Fig.~\ref{fig::tree_vs_ga}. Notably, The proposed tree-search method consistently achieves lower \ac{FoM} values with significantly fewer function evaluations, whereas \ac{GA} exhibits relatively faster improvement during the initial optimization stage. The slower convergence observed in the \ac{GA} stems from the difficulty of escaping the initial 50\% pixel occupation density, whereas the proposed tree-search algorithm is able to effectively break away from this constraint, as highlighted in Fig.~\ref{fig::pm_from_tree_and_GA}(a). Additional experiments with an increased population size of 64 show faster convergence; however, the observed performance improvement is primarily attributed to the increased number of mutation of the best-performing offspring rather than enhanced global exploration, resulting in behavior that closely resembles the proposed tree-search algorithm. Future work will investigate hybrid strategies that combine \ac{GA}-based seeding with the proposed tree-search algorithm to enable faster initialization and more efficient optimization.

\begin{figure}[t]
    \centering
    \includegraphics[width=1\columnwidth]{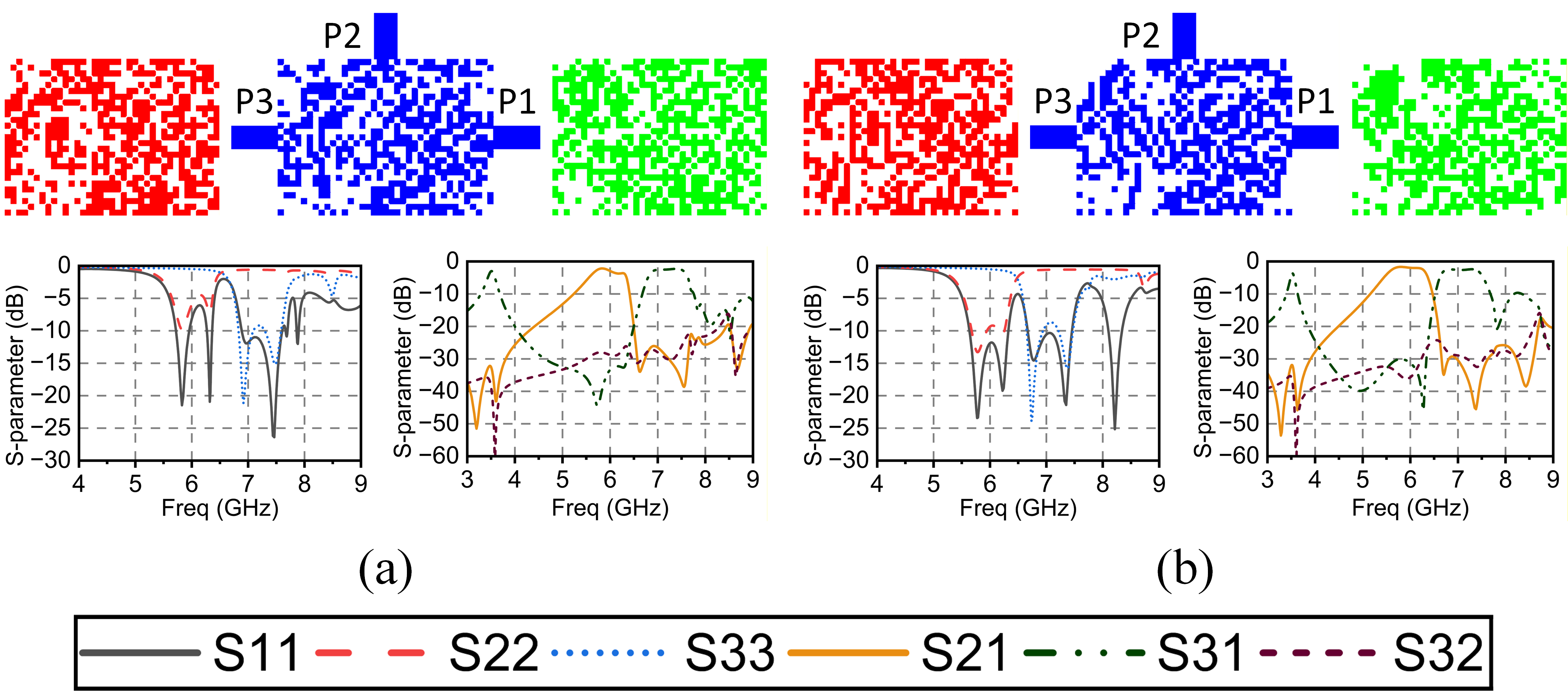}
    \caption{\textcolor{blue}{Diplexers optimized with FoM different from Eq.~(\ref{eq:rw_defs}). (a) Optimized with terms of rejection revised. (b) Optimized with terms of isolation added.}}
\label{fig::FoM_experiment}
\end{figure}

\begin{figure}[t]
    \centering
    \includegraphics[width=1\columnwidth]{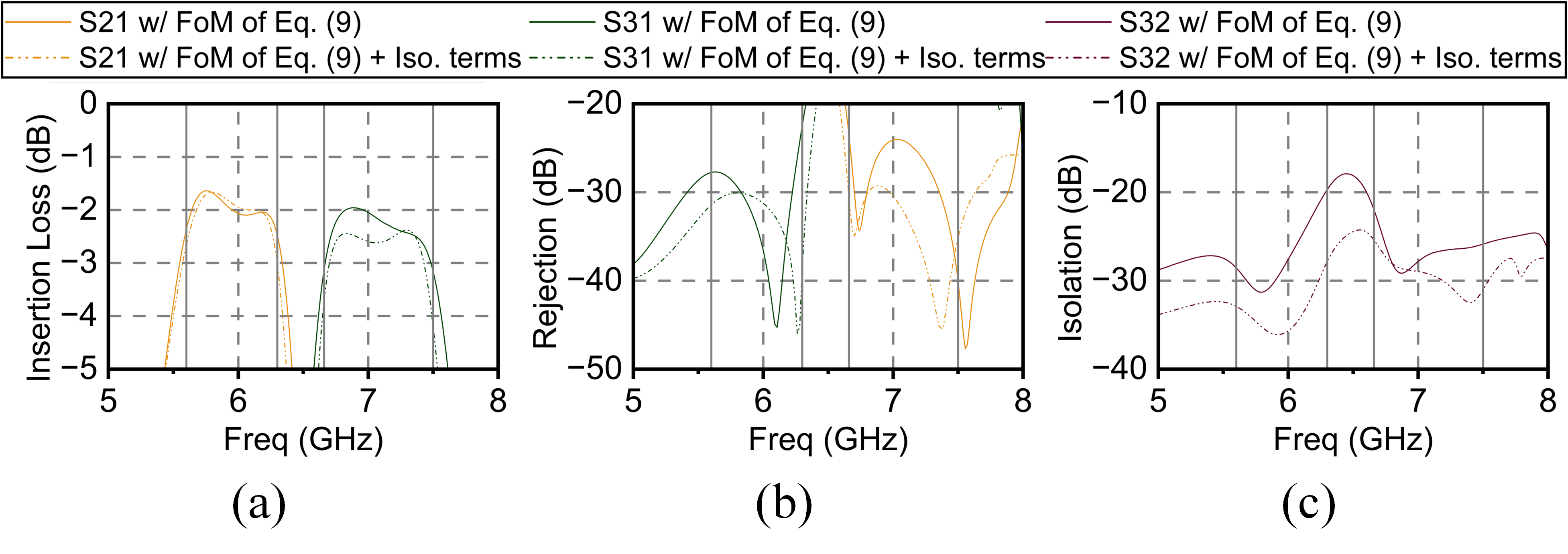}
    \caption{\textcolor{blue}{Comparison of the response of Fig.~\ref{fig::pm_from_tree_and_GA}(b) (optimized with Eq.~(\ref{eq:rw_defs}) and Fig.~\ref{fig::FoM_experiment}(b) (optimized with Eq.~(\ref{eq:rw_defs}) + isolation terms). Grey vertical lines denote the frequency boundaries in each passband. (a) Insertion loss. (b) Rejection. (c) Isolation. }}
\label{fig::FoM_experiment_comparision}
\end{figure}

\subsection{\textcolor{blue}{Impact of \ac{FoM} Weight Distribution}}
\label{subsec::fom_weight}

\textcolor{blue}{In this subsection, an impact of \ac{FoM} weight adjustment is discussed. While each coefficient in each rewarding function in Eq.~(\ref{eq:rw_defs}) is empirically adjusted, this analysis is to show how the defined \ac{FoM} guides the optimization. It is noted that the \ac{FoM} in Eq.~(\ref{eq:fom}) and Eq.~(\ref{eq:rw_defs}) is dedicated to maximize the insertion loss in each passband. The first experiment is changing the rewarding function related to the rejection, since the inter-channel attenuation is the most critical metric in the diplexer. It is noted that the diplexer with the proposed tree search algorithm with the seed of Fig.~\ref{fig::pm_from_tree_and_GA}(b). The diplexer in Fig.~\ref{fig::FoM_experiment}(a) is optimized with the rewarding functions in Eq.~(\ref{eq:rw_defs}) except $rw_{s1}(f)$ and $rw_{s2}(f)$, which is alternatively defined as below to minimize the rejection further.}

\begin{equation}
    \textcolor{blue}{rw_{s1}^{rej}(f) = 10^{\tfrac{-\Sdb{2}{1}-45}{10}},\quad rw_{s2}^{rej}(f) = 10^{\tfrac{-\Sdb{3}{1}-45}{10}}}
\end{equation}

\noindent \textcolor{blue}{However, poorer rejection in $S_{21}$ is observed in the second passband, while rejection better than $-30$~dB is achieved in the first passband. The improved rejection in the first passband is attributed to the trade-off in return loss within the same band. This additional experiment indicates that the coefficients defined in Eq.~\ref{eq:rw_defs} provide a balanced trade-off between inter-channel rejection and return loss with an emphasis on the insertion loss.}

\textcolor{blue}{An additional experiment is performed by adding additional rewarding function associated with the isolation to \ac{FoM} as below.}

\begin{equation}
    \textcolor{blue}{rw_{o_i}^{iso}(f) = 10^{\tfrac{-\Sdb{3}{2}-35}{15}}}
\end{equation}

\noindent \textcolor{blue}{This newly defined rewarding function is added up to \ac{FoM} inversely in the frequencies in $ f_{p1} \cup f_{p2}$. The optimized diplexer is shown in Fig.~\ref{fig::FoM_experiment}(b). As above, the diplexer is optimized with the seed of Fig.~\ref{fig::pm_from_tree_and_GA}(b) and the result is shown in Fig.~\ref{fig::FoM_experiment}(b). The results indicate that adding the additional terms to isolate the output ports help optimize the rejection too. For detailed comparison, Fig.~\ref{fig::FoM_experiment_comparision} is provided. Although improved isolation is expected from the modified \ac{FoM}, the inter-channel rejection also improves significantly, reaching approximately $-30$~dB in the worst case. The improvement is particularly noticeable in the second passband, with an insertion-loss trade-off of approximately 0.7~dB.}

\begin{figure*}[!tbp]
    \centering
    \includegraphics[width=0.915\textwidth]{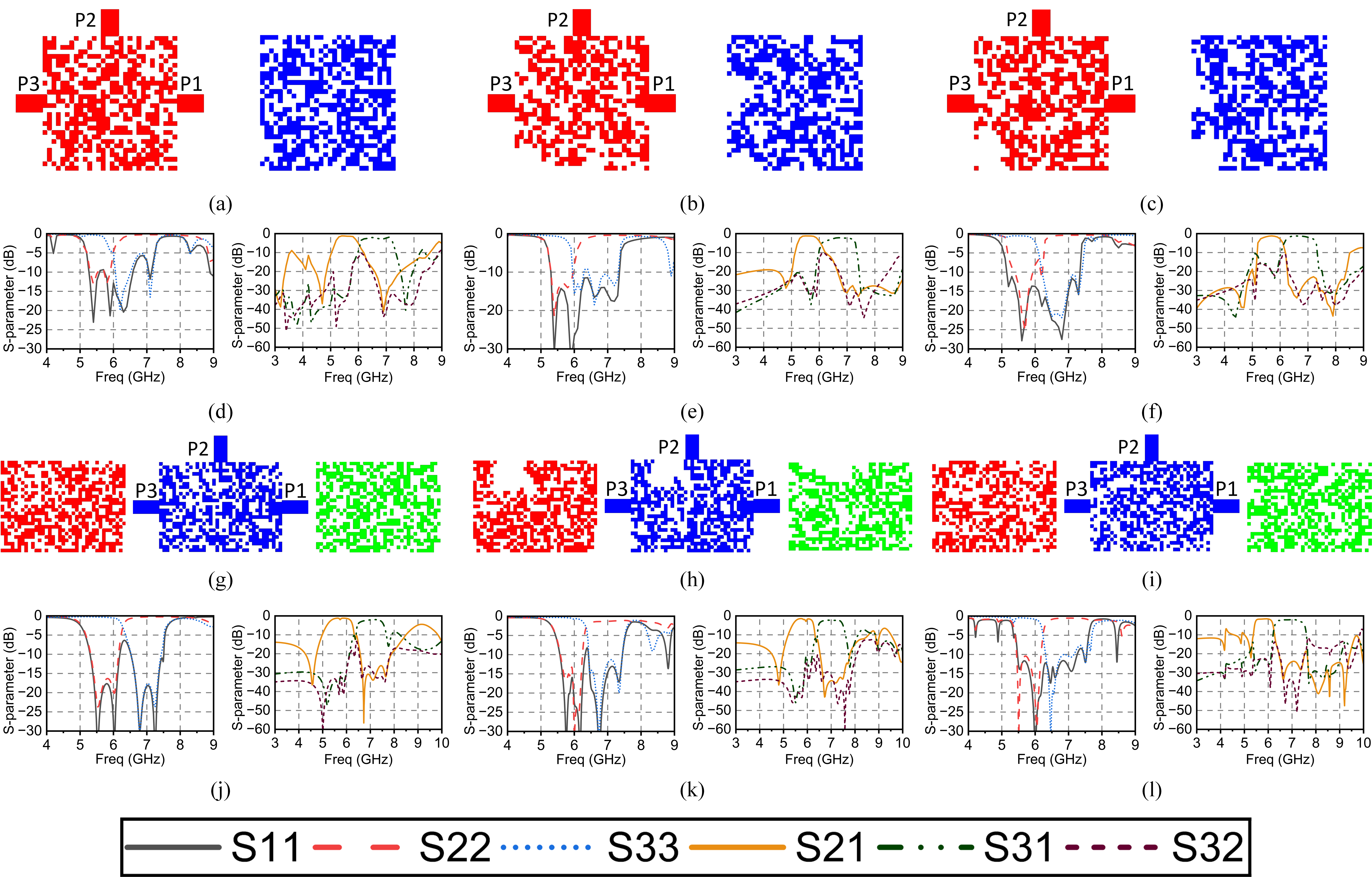}
    \caption{Optimized pixelated diplexer layouts and corresponding S-parameter responses. (a)–(c) 4-state microstrip configuration. (d)-(f) Simulation results of (a)-(c). (g)–(i) 8-state microstrip configuration. (j)-(l) Simulation results of (g)-(i).}
    \label{fig::pixel_and_s_parameter_first}
\end{figure*}

\begin{figure*}[!tbp]
    \centering
    \includegraphics[width=0.915\textwidth]{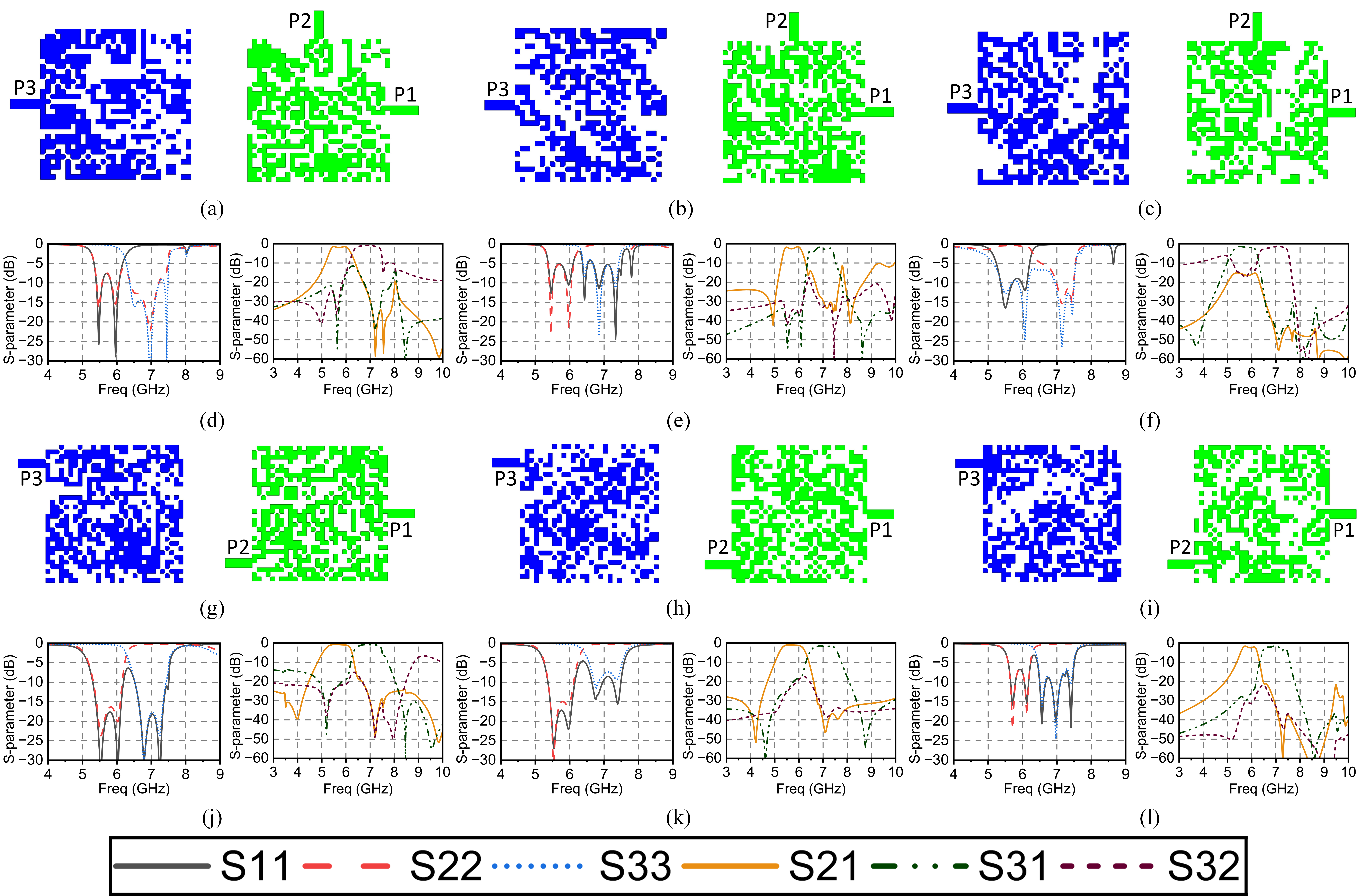}
    \caption{Optimized pixelated diplexer layouts and corresponding S-parameter responses. (a)–(c) 4-state stripline configuration. (d)-(f) Simulation results of (a)-(c). (g)–(i) 4-state stripline configuration with port positions adjusted. (j)-(l) Simulation results of (g)-(i).}
    \label{fig::pixel_and_s_parameter_second}
\end{figure*}

%% file: sec3.tex
\section{Configuration Study with Field Analysis}
\label{sec::sec3}

\begin{table}[!tbp]
\caption{Summary of Pixel Configurations and Substrate Stack-ups}
\label{tab:configuration_summary}
\centering
\resizebox{\columnwidth}{!}{%
\begin{tabular}{c|c|c|c}
\hline
\textbf{Configuration} & \textbf{Pixel Size [mm]} & \textbf{Pixel Map} & \textbf{Substrate Stack-up} \\
\hline
4 States in Microstrip & 0.5588 $\times$ 0.5588 & 30 $\times$ 30 & Fig.~\ref{fig::substrate_setup}(a): $h_1=0.25$, $h_2=0.76$ \\
8 States in Microstrip & 0.4 $\times$ 0.4 & 27 $\times$ 37 & Fig.~\ref{fig::substrate_setup}(b): $h_3=h_4=h_5=0.4866$ \\
4 States in Stripline & 0.3 $\times$ 0.3 & 30 $\times$ 30 & Fig.~\ref{fig::substrate_setup}(b): $h_3=h_4=h_5=0.4866$ \\
\hline
\end{tabular}%
}
\end{table}

In this section, the comparison across cases is carried out to examine the effects of substrate stack-up, port placement, and the number of trace layers on insertion loss, isolation, and matching. The optimized pixel configurations and corresponding S-parameter results are summarized in Fig.~\ref{fig::pixel_and_s_parameter_first} and Fig.~\ref{fig::pixel_and_s_parameter_second}. These results include three representative cases in Fig.~\ref{fig::tolerance_finite_ground}: (i) four-state microstrip, (ii) eight-state microstrip, and (iii) four-state stripline configurations. The geometrical parameters are outlined in Table \ref{tab:configuration_summary}. All simulated substrates are modeled using a homogenized Rogers~4350B dielectric with $\varepsilon_r = 3.66$ and loss tangent of 0.004. Furthermore, the input port assignment and geometrical port arrangement including its vertical and lateral position is studied in the four-state stripline configuration. 

\subsection{4 States in Microstrip}

The lateral dimensions of the pixelated surface are set to approximately 16 mm to prevent the structure from operating as a patch-like radiator at the target channels. By constraining the lateral dimension, the dominant behavior is intended to be governed by transmission-line coupling between traces, preventing the surface from entering a patch-resonant regime where cavity-like modes would complicate the optimization. The ports are distributed at three sides and the left, top, and right ports correspond to port 3, port 2, and port 1, respectively. 

\begin{figure}[!tbp]
    \centering
    \includegraphics[width=1\columnwidth]{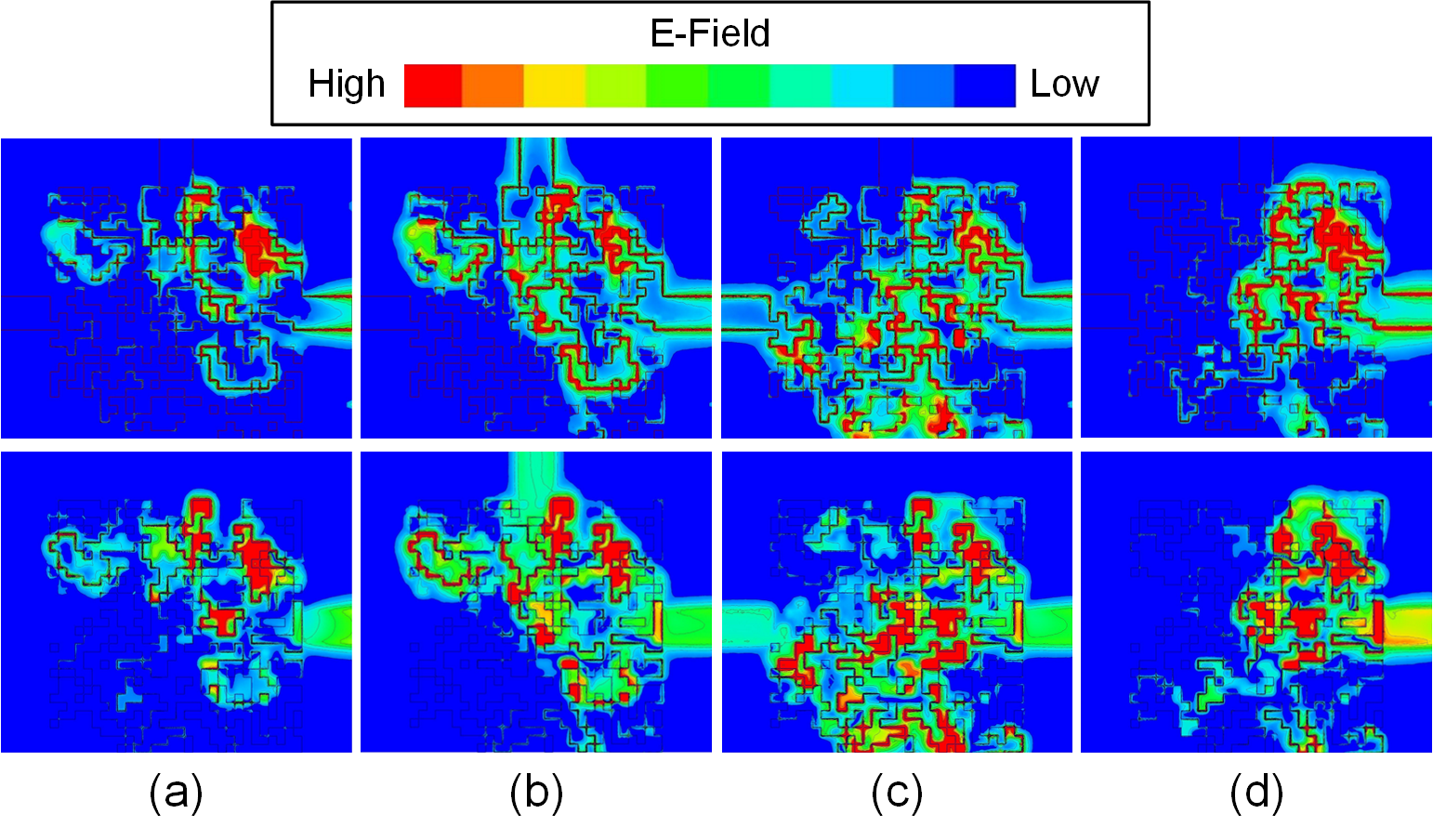}
    \caption{Complex electric field distribution at the top and bottom surfaces of Fig. \ref{fig::pixel_and_s_parameter_first}(b). (a) 4.8 GHz. (b) 5.6 GHz. (c) 6.8 GHz. (d) 7.725 GHz. }
\label{fig::fp_IMS661}
\end{figure}

Fig.~\ref{fig::pixel_and_s_parameter_first}(a)–(c) present the optimized diplexer responses. In all cases, $S_{31}$ exhibits higher insertion loss than $S_{21}$, primarily because port~3 is located farther from port~1 than port~2, resulting in a longer electrical path and increased radiation loss. While the designs do not achieve sufficient inter-channel attenuation of $-20$~dB, they maintain return losses below $-10$~dB across both passbands and achieve peak insertion losses better than $-2$~dB.

The electric-field distributions shown in Fig.~\ref{fig::fp_IMS661} correspond to Fig. \ref{fig::pixel_and_s_parameter_first}(b) when port~1 is excited, evaluated at 4.8, 5.6, 6.8, and 7.725~GHz to correlate the field evolution with the observed $S$-parameter behavior. At 5.6 GHz, corresponding to the lower passband, the field propagates toward port~2, confirming active transmission through the intended path. However, a noticeable field concentration also appears near port~3, implying unwanted lateral coupling that results in inferior $S_{31}$ rejection and slight amplitude ripple in $S_{21}$. At 7.725~GHz, in the upper stopband, the field becomes localized around the excitation ports, showing that the structure reverts to a reflective state.  Overall, these results confirm that this configuration is not yet globally optimal since none of them show enough sharp inter-channel rejections.

\begin{figure}[!tbp]
    \centering
    \includegraphics[width=1\columnwidth]{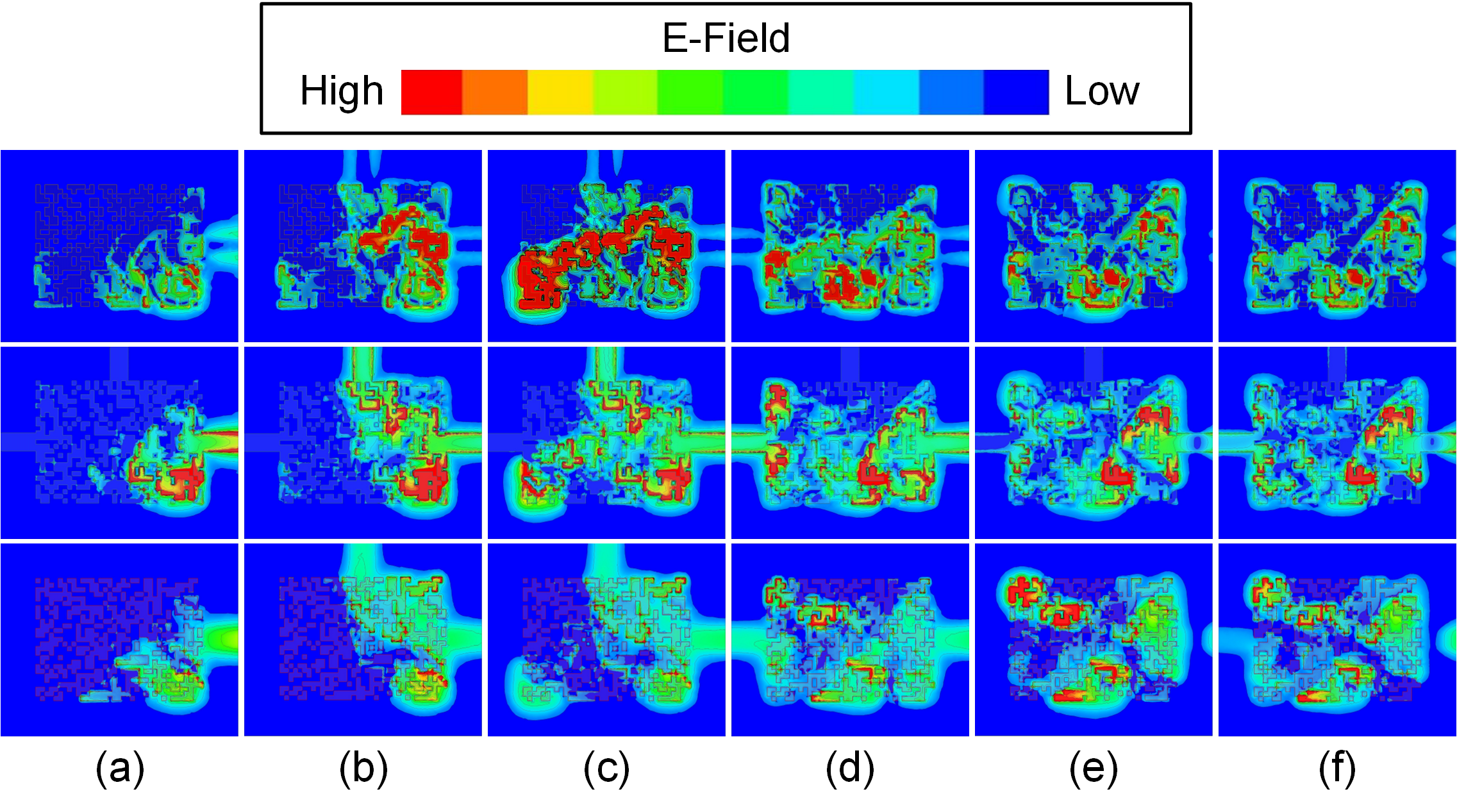}
    \caption{Complex electric field distribution at the top, the middle, and the bottom surfaces of Fig. \ref{fig::pixel_and_s_parameter_first}(g). (a) 4.5 GHz. (b) 5.4 GHz. (c) 5.6 GHz. (d) 6.7 GHz. (e) 7.7 GHz. (f) 7.9 GHz. }
\label{fig::fp_8_state_1}
\end{figure}

\begin{figure}[!tbp]
    \centering
    \includegraphics[width=1\columnwidth]{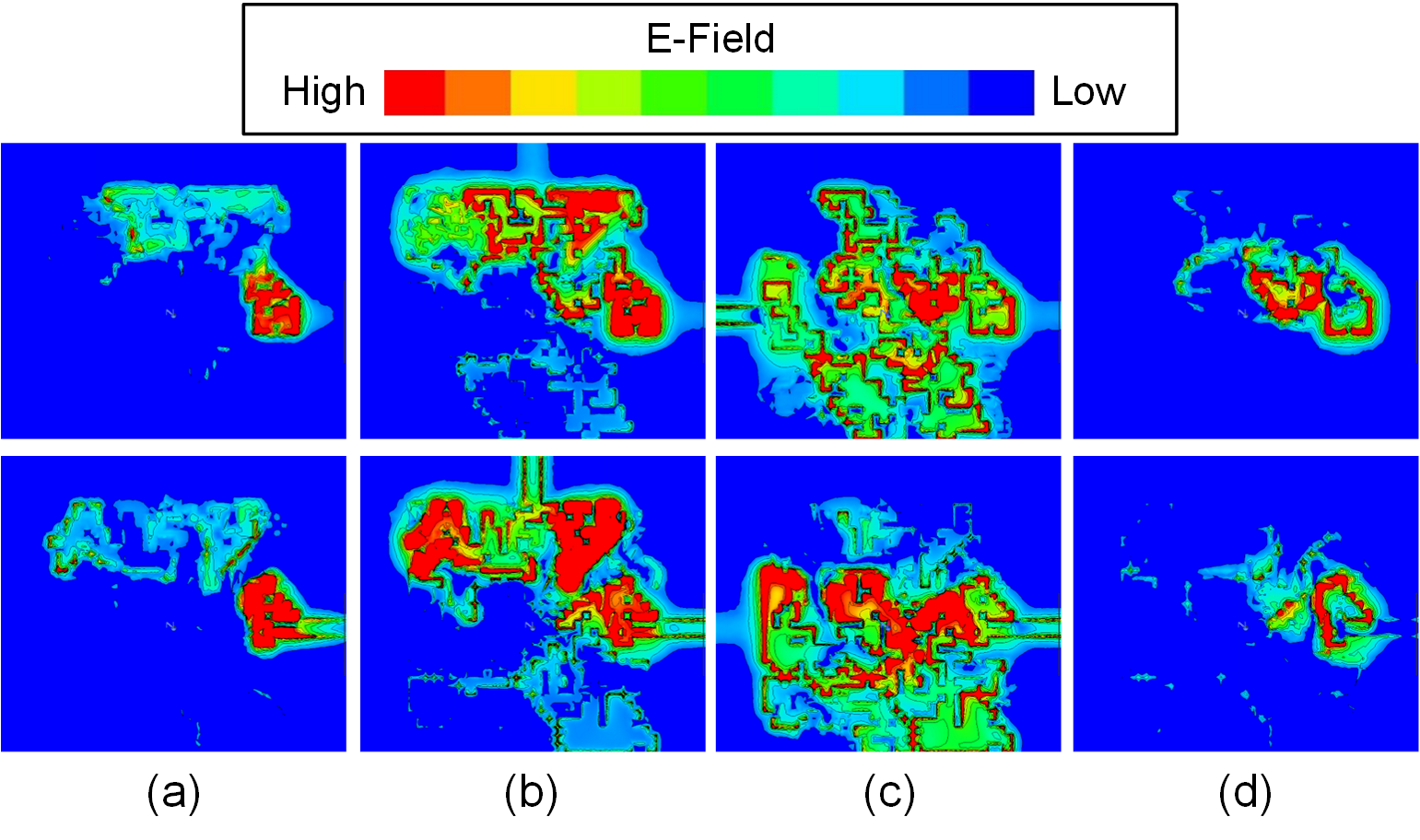}
    \caption{\textcolor{blue}{\textcolor{blue}{Complex electric field distribution at the top and bottom surfaces of Fig. \ref{fig::pixel_and_s_parameter_second}(b). (a) 4.8 GHz. (b) 5.45 GHz. (c) 6.8 GHz. (d) 8 GHz.}}}
\label{fig::fp_gv1_5}
\end{figure}

\subsection{8 States in Microstrip}

A three-layer, eight-state microstrip configuration was examined to assess the impact of additional layers on vertical coupling and inter-band isolation relative to the four-state microstrip case. The port arrangement remains the same, with the left, top, and right ports corresponding to port~3, port~2, and port~1, respectively. Lateral dimensions of 10.8mm and 14.8mm were selected to intentionally place the TM$_{10}$ and TM$_{01}$ resonances near 5 GHz and 7 GHz, respectively, in order to evaluate whether these resonant modes can be exploited for diplexer design.

As shown in Fig.~\ref{fig::pixel_and_s_parameter_first}(g)--(i), each configuration maintains the target diplexing trend with two distinct transmission bands. Overall, the frequency response is proven to show better input port matching ($S_{11}$) than the previous configuration. In particular, design~(g) demonstrates a noticable peak inter-channel rejection as much as -45 dB in channel 1 and -55 dB in channel 2. Design~(h) shows a similar response providing superior isolation ($S_{32}$ under -30~dB across both two bands and the intermediate band). Overall, the three-layer configuration offers improved performance but only spectral benefits relative to the four-state case in spite of the increased number of states.

The field distributions in Fig.~\ref{fig::fp_8_state_1}, corresponding to port~1 excitation, provide additional insight into these spectral characteristics. At 4.5~GHz, the field remains confined near the feed, representing strong reflection and negligible coupling to the other ports. At 5.4~GHz, the field begins to propagate toward port~2, marking the onset of the first passband. To inspect the slight notch at the pass-band shown in Fig.~\ref{fig::pixel_and_s_parameter_first}(j), the field distributions at 5.6 GHz is performed. A pronounced lateral electromagnetic resonance appears across the top layer, resembling a $TM_{10}$ mode that adds a parasitic path from port 1 to port 3 and explains this shallow notch observed in $S_{21}$. This behavior suggests that it is hard to make harmonious interplay between transmission-line based response and resonator-based response. At 6.7~GHz, within the second passband, the field couples effectively from port~1 to port~3 through the middle layer without parasitic peaks. At 7.7~GHz, the energy becomes trapped away from both port~1 and port~3 with multiple electric field peaks, corresponding to rejection of -55 dB; in contrast, at 7.9~GHz, unwanted field propagation emerges along the bottom side, indicating that such combination of electric field peaks are inherently narrow-band.

Nevertheless, parasitic resonances are more pronounced throughout the designs than in the designs in the previous microstrip 4-state configurations. This indicates that it is difficult to achieve harmonious interplay between resonator-based responses and responses based on local coupling. Furthermore, it is observed that the optimized pixel maps in Fig.~\ref{fig::pixel_and_s_parameter_first}(h) exhibit vacant regions near the upper edges of the middle and top layers, which contribute to improved isolation between port~2 and port~3. However, the presence of these unused regions suggests that the left–top–right port placement might be suboptimal, as portions of the available design space are not effectively utilized. Motivated by this observation, the impact of alternative port placements is investigated in the following subsection.

\textcolor{blue}{Among the examined designs, the diplexer shown in Fig.~\ref{fig::pm_from_tree_and_GA}(a) achieves the most balanced diplexing behavior with moderate inter-channel attenuation, exceeding -20~dB across both operating bands, while exhibiting a notably reduced occupation density in the middle layer and strategic patterning to suppress $TM_{10}$ mode. The diplexer in ~\ref{fig::FoM_experiment}(b) shows the most pronounced inter-channel attenuation, exceeding -30~dB across both operating bands, although trading off insertion loss and return loss. These results indicate that the eight-state configuration has the potential of improving inter-channel rejection as it offers richer signal propagation/trapping paths; nevertheless, further refinement of the optimization strategy is required to achieve precise resonance control across all layers.}

\begin{figure}[tbp!]
    \centering
    \includegraphics[width=1\columnwidth]{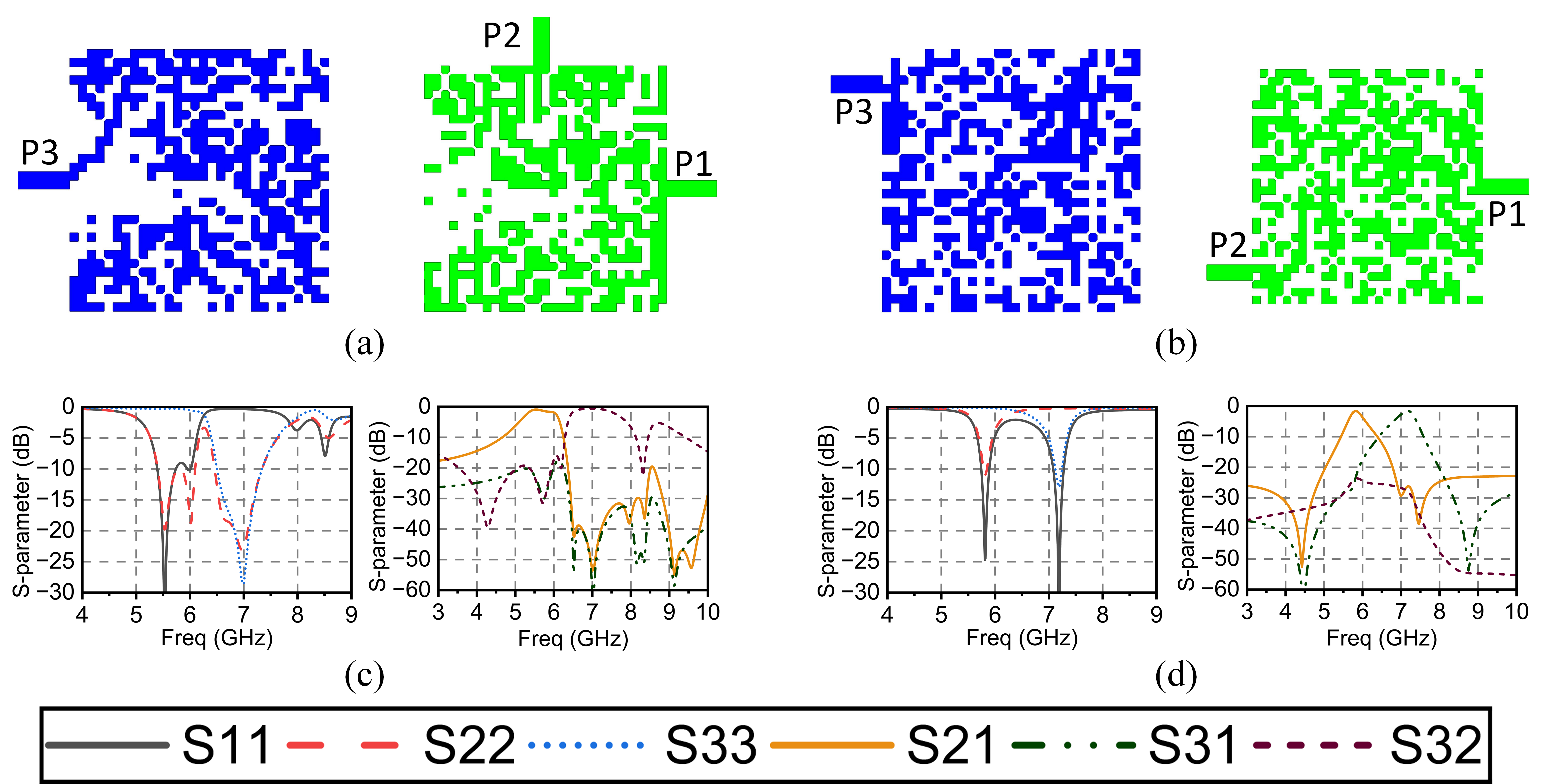}
    \caption{Additional pixel maps with their frequency responses.} 
\label{fig::additional_pixelmap}
\end{figure}

\begin{figure}[tbp!]
    \centering
    \includegraphics[width=1\columnwidth]{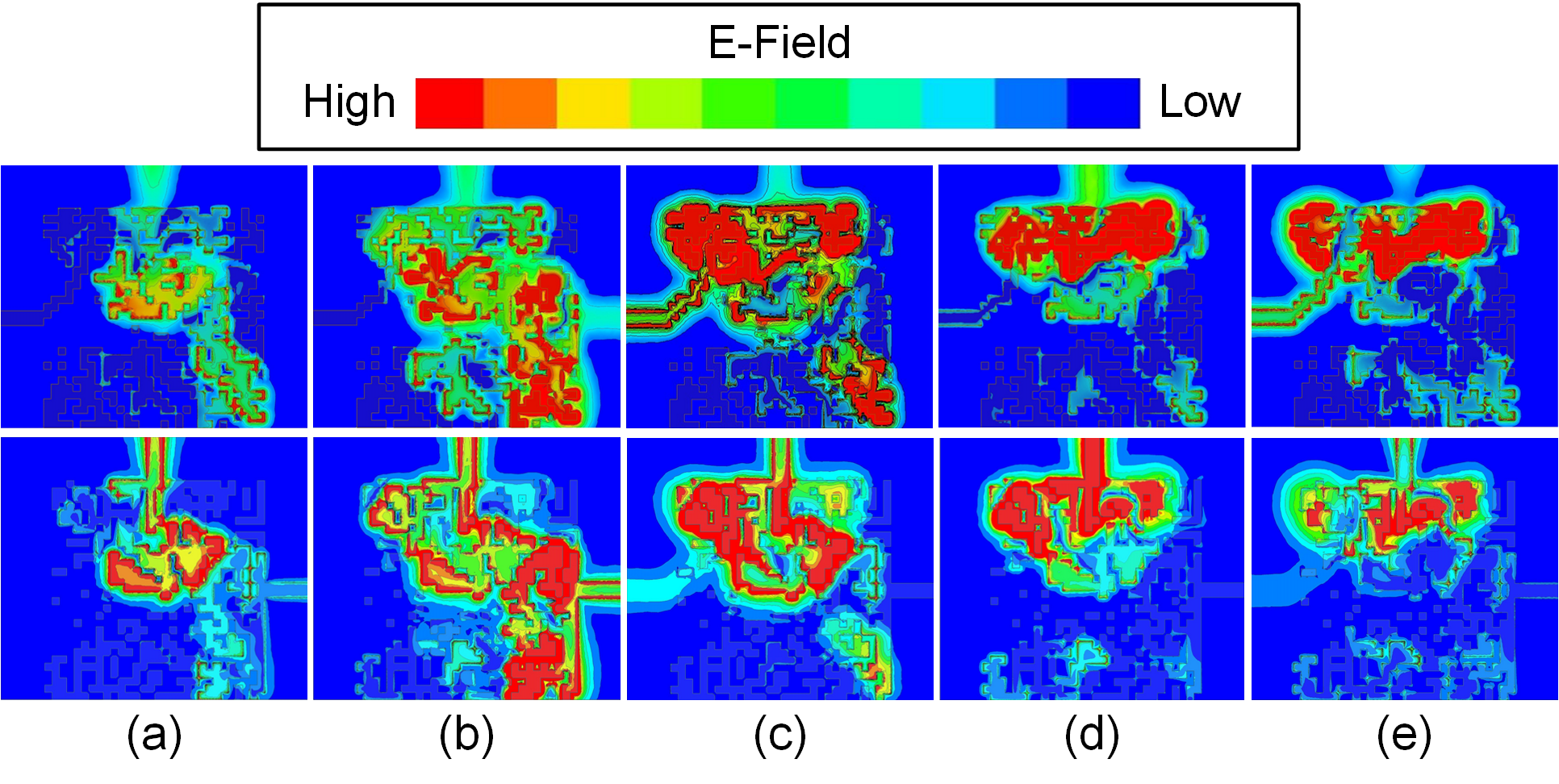}
    \caption{Complex electric field distribution at the top and bottom surfaces of Fig. \ref{fig::additional_pixelmap}(a). (a) 4.3 GHz. (b) 5.7 GHz. (c) 6.5 GHz. (d) 8.2 GHz. (e) 8.6 GHz.}
\label{fig::fp_gv1_2}
\end{figure}

\begin{figure}[tbp!]
    \centering
    \includegraphics[width=1\columnwidth]{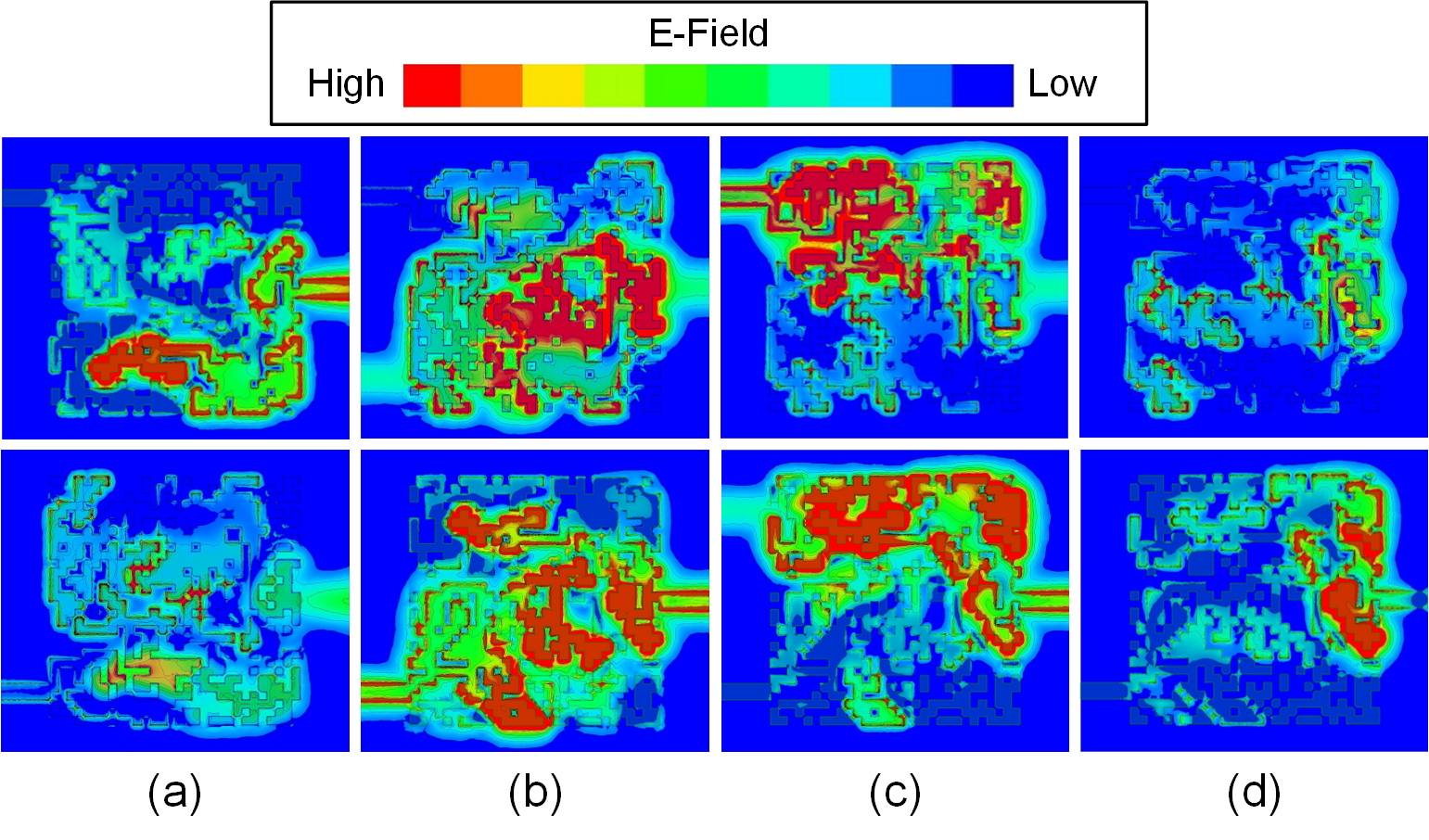}
    \caption{Complex electric field distribution at the top and bottom surfaces of Fig. \ref{fig::pixel_and_s_parameter_second}(g). (a) 4 GHz. (b) 5.7 GHz. (c) 7.15 GHz. (d) 8.5 GHz.}
\label{fig::fp_gv2_1}
\end{figure}

\subsection{4 States in Stripline with Different Port Position}

The four-state stripline configuration was developed as an alternative to the eight-state design, which exhibited excessive optimization complexity and limited practical improvement. To suppress radiation loss, both the top and bottom sides of the substrate are grounded, forming a fully enclosed stripline environment that confines electromagnetic energy within the dielectric region. The total lateral dimension was also reduced to 9~mm to prevent $TM_{10}$ or $TM_{01}$ resonance at both two pass bands. 

Two port arrangements were evaluated to study the effect of port placement on frequency response and field behavior. In Fig.~\ref{fig::pixel_and_s_parameter_second}(a)–(c), the three ports are positioned on three different edges, where the left, top, and right edges correspond to port~3, port~2, and port~1, respectively. In contrast, Fig.~\ref{fig::pixel_and_s_parameter_second}(g)–(i) employs a configuration with one input port on one side and two output ports on the opposite side located on different metal layers, where the left–top, left–bottom, and right ports correspond to port~3, port~2, and port~1, respectively. The latter configuration is later shown to provide better performance, as the field needs not to be inverted vertically. Moreover, placing the excitation ports on different layers introduces an additional roll-off behavior, which facilitates the optimization process.

\textcolor{blue}{The designs in Fig. \ref{fig::pixel_and_s_parameter_second}(a)–(c) each show at least one weakness—typically insufficient out-of-band rejection or degraded return loss. It is noted that the common output port of the diplexer varies from (a) to (c), each port 2, port 1, and port 3. Despite extensive variation of port configurations, none of the optimized designs achieves a fully balanced diplexing response. Fig.~\ref{fig::fp_gv1_5} shows the field distributions of the diplexer in Fig.~\ref{fig::pixel_and_s_parameter_second}(b), which exhibits the most balanced behavior, albeit with degraded return loss. Fig.~\ref{fig::fp_gv1_5}(a) and (b) indicate that multilayer traces can both create notches through field cancellation at the input port and support passband transmission (from port 1 to port 2) at other frequencies. Fig.~\ref{fig::fp_gv1_5}(c) shows that while the structure enables the field propagation toward port~3, it does not fully suppress coupling to port~2 along the top traces.} 

\textcolor{blue}{The field distributions in Fig.~\ref{fig::fp_gv1_2} illustrate the behavior of the configuration in Fig.~\ref{fig::additional_pixelmap}(a), with the common output port as port 2. It is noted that this field is excited with the port~2. This case is selected because it demonstrates strong roll-off in both $S_{21}$ and $S_{31}$ at intermediate frequencies. At 5.7~GHz and 6.5~GHz, the diplexing behavior emerges clearly, indicating effective channel separation. Fig.~\ref{fig::fp_gv1_2}(c) is worth noting since additional lower-right region of the top layer helps reject the transmission through port~1, resulting in the sharp roll-off. At 8.6~GHz [Fig.~\ref{fig::fp_gv1_2}(e)], additional leakage between the two output ports appears, confirming that this port placement is not ideal at higher frequencies due to close port placements. Overall, this case verifies that placing three ports on three separate sides of the pixelated surface can achieve diplexing behavior, but the spacing between the output ports still needs more optimization.}

The designs in Fig. \ref{fig::pixel_and_s_parameter_second}(g)–(i) demonstrate the most balanced diplexing behavior, which indicates that the output ports at the same side but different layers are effective to introduce diplexing response. The design (i) reveals that a conventional 50~$\Omega$ termination may not correspond to the optimum impedance condition for this topology. Although its reflection coefficient is degraded, the resulting responses show an exceptionally high roll-off, highlighting the possibility that setting the port impedance as an alternative complex impedance can be the optimum port impedance for certain configurations. Fig.~\ref{fig::additional_pixelmap}(b) shows that the same optimization framework can also produce a narrowband diplexer response. This case corresponds to an intermediate stage of the all-way tree-search process, indicating that the algorithm is capable of identifying high-Q responses first and then expand the bandwidth during subsequent iterations if guided well.

Fig.~\ref{fig::fp_gv2_1} examines the configuration in Fig.~\ref{fig::pixel_and_s_parameter_second}(g), which exhibits the most balanced diplexing behavior across the band. Between 5.7~GHz and 7.15~GHz, two distinct field-guided paths form for the two operating bands. The rejection at 4 GHz and 8.5 GHz is also notable since it fully exploits the lateral traces and vertical traces to suppress the transmission. These results indicate that larger lateral spacing between the input port and the output ports, different from the port configuration in Fig. \ref{fig::pixel_and_s_parameter_second}(a)–(c), yields a more balanced diplexing response. Although this configuration shows the most balanced diplexing behavior, identifying an optimal port placement with geometrical configurations still requires more exhaustive exploration, as well as different spacing and orientation combinations. Sub-optimal response is also not irrelevant to the search process and the definition of \ac{FoM}.

%% file: sec4.tex
\section{Meshing Strategy with Computation Cost}
\label{sec::sec4}

In this section, meshing strategy to find balance speed and accuracy in \ac{MoM} solver with the proposed pre-computation scheme is discussed.

\subsection{Number of Mesh and Orientation Distribution}
\label{subsection::number_of_mesh}

\begin{figure}[t]
    \centering
    \includegraphics[width=1\columnwidth]{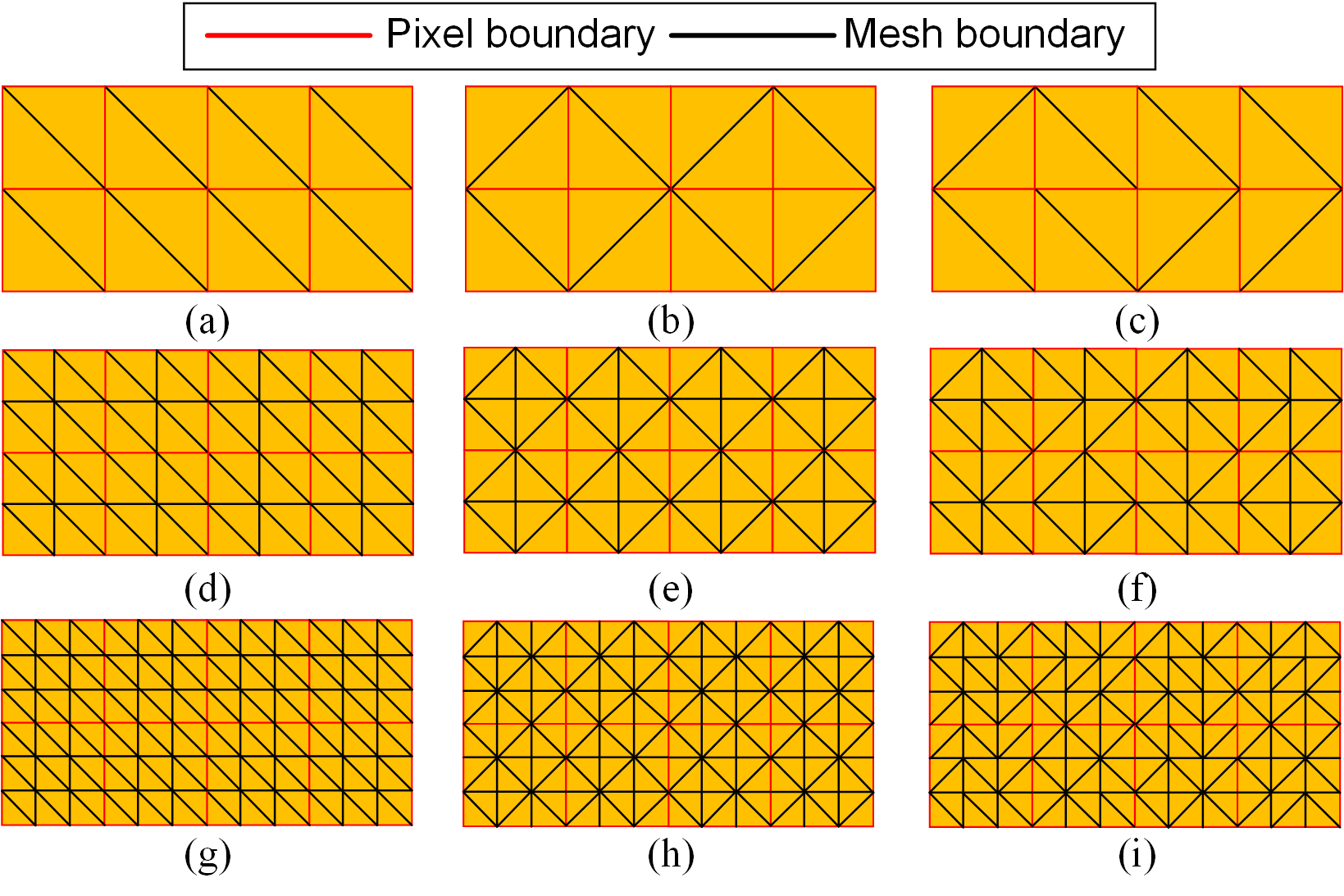}
    \caption{Mesh orientation configurations. (a) 2 uniformly oriented triangles per pixel. (b) 2 alternatingly oriented triangles per pixel. (c) 2 randomly oriented triangles per pixel. (d) 8 uniformly oriented triangles per pixel. (e) 8 alternatingly oriented triangles per pixel. (f) 8 randomly oriented triangles per pixel. (g) 18 uniformly oriented triangles per pixel. (h) 18 alternatingly oriented triangles per pixel. (i) 18 randomly oriented triangles per pixel.}
\label{fig::mesh_orientation}
\end{figure}

While the \ac{RWG} basis function with basic two-triangle meshing has been used in prior studies \cite{gamom}, no systematic analysis has been conducted on how mesh density influences the accuracy of the \ac{MoM} relative to a full-wave solver. To address this, part of the “parent” structure used for generating the initial interaction matrix was analyzed with varying mesh densities, as shown in Fig.~\ref{fig::mesh_orientation}, where each pixel is subdivided into 2, 8, or 18 right triangles. It should be noted that the number of triangles does not directly correspond to the number of \ac{BF}, since the \ac{RWG} \ac{BF} formulation defines basis functions over shared triangle edges rather than per element.

\begin{figure}[t]
    \centering
    \includegraphics[width=1\columnwidth]{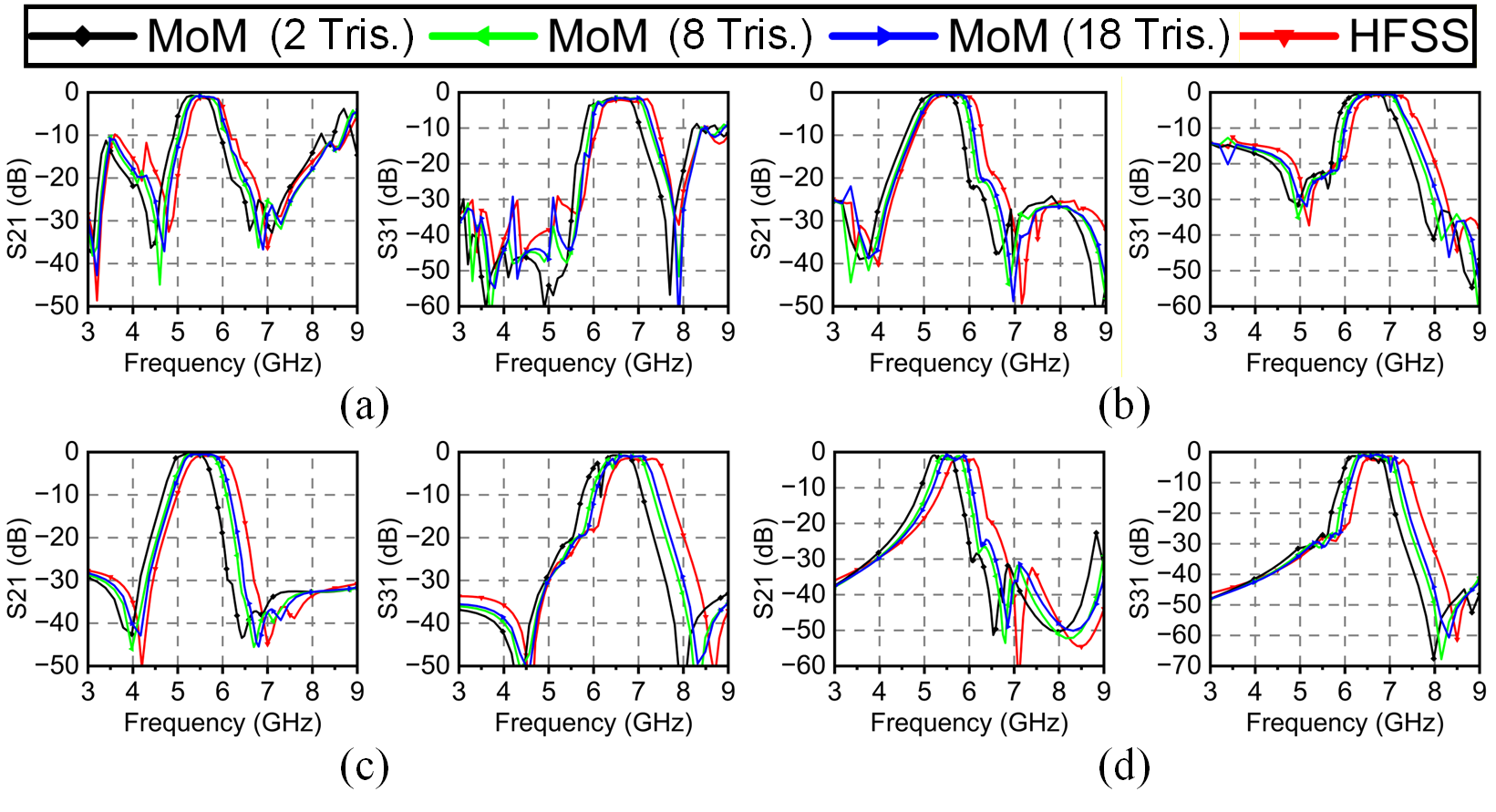}
    \caption{Comparison of frequency responses obtained from the \ac{MoM}-based solver with varying numbers of triangular elements per the side in a pixel \textcolor{blue}{with the diagonal distribution fixed to alternating case} against results from Ansys HFSS. (a) Fig. \ref{fig::pixel_and_s_parameter_first}(a). (b) Fig. \ref{fig::pixel_and_s_parameter_second}(g). (c) Fig. \ref{fig::pixel_and_s_parameter_second}(h). (d) Fig. \ref{fig::pixel_and_s_parameter_second}(i).}
\label{fig::frequency_shift_graph}
\end{figure}

Consequently, one of the optimized filters was selected to perform a parametric study on mesh density versus S-parameter accuracy in terms of \ac{RMSE}. The accuracy of the \ac{MoM} solver is evaluated by varying the number of triangular elements used to discretize each pixel. These configurations represent different trade-offs between computational cost and electromagnetic accuracy. The S-parameter comparisons with HFSS are presented in Fig.~\ref{fig::frequency_shift_graph}, showing that finer meshes yield improved agreement with the full-wave results. As shown in Fig.~\ref{fig::mesh_orientation}, the number of triangular elements per pixel was varied from 2 to 8 and 18, while much denser 32-triangle discretization was not pursued due to its excessive computational requirements. The frequency responses indicate a gradual convergence toward the HFSS reference as the mesh density increases. However, slight frequency shifts can still be observed in all cases. In particular, the coarsest mesh \textcolor{blue}{overestimates} the effective electrical length, causing a leftward shift in resonant frequencies, whereas finer meshes gradually right-shift the response to align with the full-wave solver.

Three mesh orientation distributions were examined: uniform diagonal, alternating diagonal, and randomized diagonal patterns, as shown in Fig.~\ref{fig::mesh_orientation}. At first glance, the alternating orientation can be regarded as the most uniform mesh distribution, as it provides consistent current representation both within and between adjacent pixels by alternating the diagonal direction. This configuration improves consistency compared to the uniformly oriented case, where all RWG basis functions are aligned in the same direction and consequently produce uneven current sharing across neighboring pixels. The alternating distribution maintains geometric symmetry and makes the current distribution representation more even. The randomly oriented configuration, on the other hand, introduces stochastic variations in current distribution relative to the true current profile, which can cause fluctuations in the S-parameter across different pixel maps. Therefore, the uniformly diagonal case seems to be the least effective configuration, since it lacks both the systematic uniformity of the alternating case and the \textcolor{blue}{potentially} beneficial variability of the random case. Another meshing strategy can involve non-uniform triangularization, such as using finer elements near edges \cite{woojun_APMC}. \textcolor{blue}{Section~\ref{subsection::rmse_method_and_results} will identify the mesh configuration that finds balance between the optimization speed and accuracy per configuration.}

\subsection{Optimization Speed-up}
\label{subsection::computational_cost}
In this subsection, time consumed to simulate a single S-parameter of each design is provided. The computational cost of the proposed optimization framework can be categorized into two primary components: (1) the pre-computation of impedance matrix elements in the \ac{MoM} solver, and (2) the optimization time using the depth-increasing all-way tree navigation search algorithm, where the time of the matrix inversion time dominates. The matrix pre-computation was executed on a 128-core AMD EPYC~7702 processor, where 72 frequency-dependent matrices were generated. For example, the pre-computation of the 72 matrices for the stripline four-state configuration with eight triangles per pixel required approximately 4 hours.

\begin{table}[t]
\caption{Computational Cost and Optimization Statistics}
\label{tab:computational_cost}
\centering
\resizebox{\columnwidth}{!}{%
\begin{tabular}{c|c|c}
\hline
\textbf{Metric/Config.} & \textbf{Fig.~\ref{fig::tolerance_finite_ground} (b)} & \textbf{Fig.~\ref{fig::tolerance_finite_ground} (c)} \\
\hline
Parent matrix dimension 
& \begin{tabular}[c]{@{}c@{}}
2 tris: $9824 \times 9824$ \\
8 tris: $36588 \times 36588$
\end{tabular}
& \begin{tabular}[c]{@{}c@{}}
8 tris: $21865 \times 21865$
\end{tabular} \\
\hline
Averaged matrix inversion time 
& \begin{tabular}[c]{@{}c@{}}
2 tris: 0.35 s (12 L40S GPUs) \\
8 tris: 3.5 s (8 H200 GPUs)
\end{tabular}
& \begin{tabular}[c]{@{}c@{}}
8 tris: 0.95 s (12 L40S GPUs)
\end{tabular} \\
\hline
Average number of inversion
& 8500 (max. depth = 2)
& 26000 (max. depth = 4) \\
\hline
Average total optimization time
& \begin{tabular}[c]{@{}c@{}}
2 tris: 50 min \\
8 tris: 8 h 17 min
\end{tabular}
& \begin{tabular}[c]{@{}c@{}}
8 tris: 6 h 51 min
\end{tabular} \\
\hline
\end{tabular}%
}
\end{table}

The computational cost during the optimization process is summarized in Table~\ref{tab:computational_cost}. For Fig.~\ref{fig::tolerance_finite_ground}(b), both two-triangle and eight-triangle per-pixel meshing strategies were evaluated, whereas Fig.~\ref{fig::tolerance_finite_ground}(c) employed only the eight-triangle configuration, as its 4-state configuration results in a reduced computational burden and less numbers of flipping direction. To mitigate the cubic scaling of computational cost associated with matrix inversion, the proposed framework employs GPU acceleration, which significantly reduces the overall runtime during the iterative inverse design process. The matrix inversion tasks were offloaded to clusters consisting of either 12 NVIDIA~L40S GPUs (48~GB each), 36 NVIDIA~A30 GPUs (24~GB each), or 8 NVIDIA~H200 GPUs (141~GB each). utilizing \ac{MPI}-based distributed parallelization for acceleration. Overall, the complete optimization for each configuration was completed within one day, demonstrating the scalability and computational efficiency of the proposed \ac{MPI}- and \ac{GPU}-accelerated \ac{MoM} framework, with additional refinement possible given extended runtime.

While a full-wave HFSS simulation of the optimized diplexers in Fig.~\ref{fig::tolerance_finite_ground}(b) requires approximately 17~hours and those in Fig.~\ref{fig::tolerance_finite_ground}(c) require about 13~hours, the proposed approach yields an apparent computational acceleration on the order of $10^{5}$. A direct comparison with commercial solvers coupled with global optimizers is impractical due to the prohibitive computational cost. However, this speedup is achieved at the expense of trading off accuracy due to the reduced mesh density used to model the structures. Consequently, an \ac{RMSE} analysis of each S-parameter is conducted in the following subsections to quantify the impact of mesh orientation and mesh density on the simulation accuracy.

\subsection{RMSE Analysis Method and Results}
\label{subsection::rmse_method_and_results}
In this subsection, the accuracy compared with the commerical solver Ansys HFSS with respect to the mesh configuration in Fig.~\ref{fig::mesh_orientation} is extensively analyzed \textcolor{blue}{to determine the best meshing configuration to find balance between optimization speed and accuracy.} As shown in Fig.~\ref{fig::frequency_shift_graph}, frequency shift is consistently observed between the \ac{MoM} solver and Ansys HFSS simulations throughout the different configurations. \textcolor{blue}{To compensate this, \ac{DTW}–based frequency alignment method was implemented in MATLAB, which is integrated into the optimization loop. It is noted that this frequency warp is global within the same configuration. For example, the designs in Fig.~\ref{fig::pixel_and_s_parameter_second}(a)-(c) and  Fig.~\ref{fig::pixel_and_s_parameter_second}(g)-(i) share the same frequency warp.} The \ac{DTW} algorithm, originally proposed in \cite{dtw}, is a dynamic programming approach that measures similarity between two temporal sequences by nonlinearly aligning them in the frequency domain, thereby compensating for local frequency deviations. \textcolor{blue}{In this work, all the entry in [S] matrix was chosen as the alignment anchors.} After alignment, the \ac{RMSE} is computed in logarithmic space to quantify the error. For fair comparison, the randomized mesh configuration was fixed for each triangle at a designated position across all tested pixel geometries within the same configuration by setting a constant seed value in a random number generator. In this work, MATLAB implementation of DTW~\cite{MATLAB} was used for this analysis.

\renewcommand\cellalign{cc}
\renewcommand\theadalign{cc}

\begin{table}[t]
\caption{\textcolor{blue}{RMSE across different mesh configurations}}
\label{tab:rmse_with_no_of_mesh}
\centering
\resizebox{\columnwidth}{!}{%
\begin{tabular}{l|ccc|ccc|ccc|ccc}
\hline
\multirow{2}{*}{\makecell[c]{\textbf{Design/}\\\textbf{No. of Tris.}}}
& \multicolumn{3}{c|}{\textbf{S11}}
& \multicolumn{3}{c|}{\textbf{S21}}
& \multicolumn{3}{c|}{\textbf{S31}}
& \multicolumn{3}{c}{\textbf{S22}} \\
\cline{2-13}
& \textbf{2-a} & \textbf{8-a} & \textbf{18-a}
& \textbf{2-a} & \textbf{8-a} & \textbf{18-a}
& \textbf{2-a} & \textbf{8-a} & \textbf{18-a}
& \textbf{2-a} & \textbf{8-a} & \textbf{18-a} \\
\hline
Fig. \ref{fig::pixel_and_s_parameter_second}(g)
& 4.50 & 2.61 & 2.91
& 2.64 & 1.79 & 1.91
& 2.53 & 2.07 & 2.36
& 1.82 & 2.38 & 2.02 \\
Fig. \ref{fig::pixel_and_s_parameter_second}(h)
& 4.06 & 3.03 & 2.07
& 3.24 & 2.68 & 2.54
& 6.53 & 4.68 & 3.91
& 2.61 & 2.14 & 1.71 \\
Fig. \ref{fig::pixel_and_s_parameter_second}(i)
& 3.90 & 3.32 & 2.41
& 4.66 & 3.74 & 4.01
& 3.94 & 3.69 & 3.15
& 3.95 & 3.19 & 1.78 \\
Fig. \ref{fig::pixel_and_s_parameter_first}(a)
& 2.62 & 2.04 & 1.36
& 3.35 & 3.64 & 2.56
& 8.74 & 6.27 & 5.79
& 0.76 & 0.69 & 0.55 \\
\hline
\multicolumn{13}{c}{} \\[-1.0ex]
\hline
\multirow{2}{*}{\makecell[c]{\textbf{Design/}\\\textbf{No. of Tris.}}}
& \multicolumn{3}{c|}{\textbf{S23}}
& \multicolumn{3}{c|}{\textbf{S33}}
& \multicolumn{6}{c}{\textbf{Global}} \\
\cline{2-13}
& \textbf{2-a} & \textbf{8-a} & \textbf{18-a}
& \textbf{2-a} & \textbf{8-a} & \textbf{18-a}
& \multicolumn{2}{c}{\textbf{2-a}}
& \multicolumn{2}{c}{\textbf{8-a}}
& \multicolumn{2}{c}{\textbf{18-a}} \\
\hline
Fig. \ref{fig::pixel_and_s_parameter_second}(g)
& 3.67 & 2.08 & 2.50
& 2.86 & 1.86 & 1.57
& \multicolumn{2}{c}{3.01}
& \multicolumn{2}{c}{2.15}
& \multicolumn{2}{c}{2.25} \\
Fig. \ref{fig::pixel_and_s_parameter_second}(h)
& 2.82 & 1.57 & 1.39
& 2.08 & 1.52 & 0.99
& \multicolumn{2}{c}{3.85}
& \multicolumn{2}{c}{2.82}
& \multicolumn{2}{c}{2.31} \\
Fig. \ref{fig::pixel_and_s_parameter_second}(i)
& 3.50 & 2.68 & 2.45
& 1.46 & 1.88 & 2.05
& \multicolumn{2}{c}{3.71}
& \multicolumn{2}{c}{3.15}
& \multicolumn{2}{c}{2.75} \\
Fig. \ref{fig::pixel_and_s_parameter_first}(a)
& 5.84 & 5.03 & 4.18
& 2.03 & 1.66 & 1.30
& \multicolumn{2}{c}{7.36}
& \multicolumn{2}{c}{3.77}
& \multicolumn{2}{c}{3.20} \\
\hline
\end{tabular}%
}
\end{table}

\begin{figure}[!t]
    \centering
    \includegraphics[width=1\columnwidth]{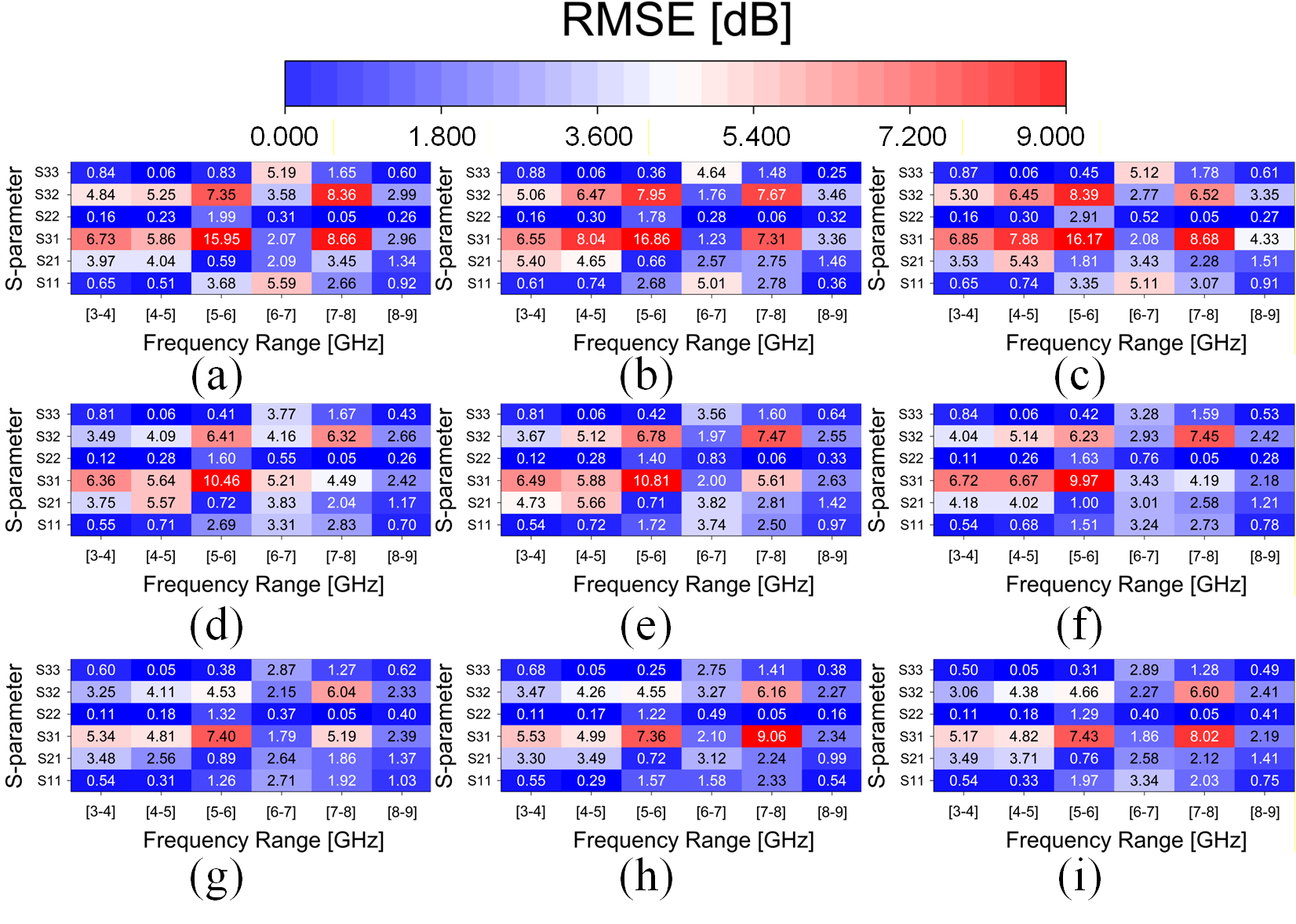}
    \caption{\textcolor{blue}{\ac{RMSE} of S-parameters against ANSYS HFSS for Fig.~\ref{fig::pixel_and_s_parameter_first}(a) under the nine mesh configurations of Fig.~\ref{fig::mesh_orientation}. Each \ac{RMSE} in (a)–(i) correspond to Fig.~\ref{fig::mesh_orientation}(a)–(i).}}
\label{fig::rms_IMS_363}
\end{figure}

\begin{figure}[!t]
    \centering
    \includegraphics[width=1\columnwidth]{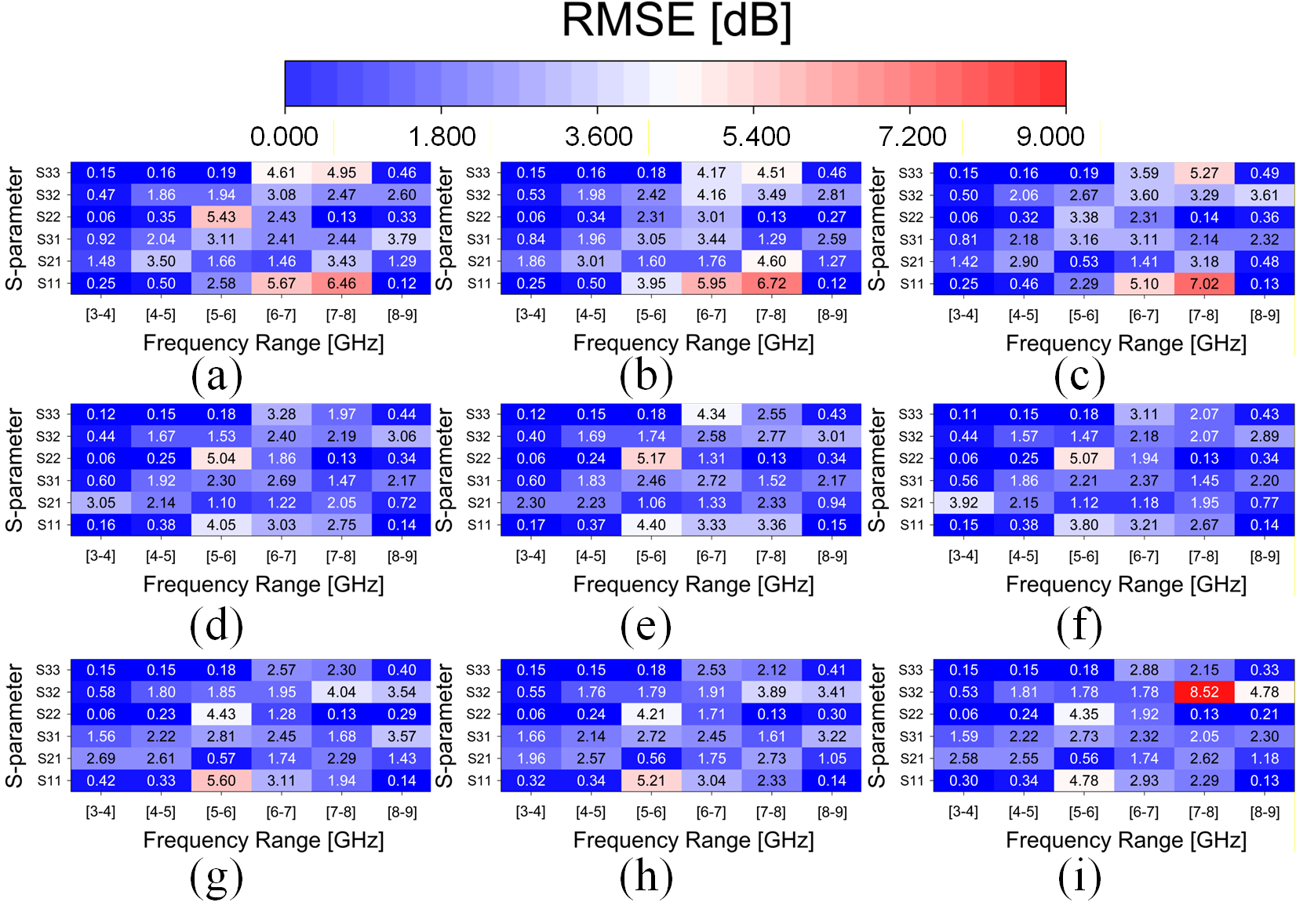}
    \caption{\textcolor{blue}{\ac{RMSE} of S-parameters against ANSYS HFSS for Fig.~\ref{fig::pixel_and_s_parameter_second}(g) under the nine mesh configurations of Fig.~\ref{fig::mesh_orientation}. Each \ac{RMSE} in (a)–(i) correspond to Fig.~\ref{fig::mesh_orientation}(a)–(i).}}
\label{fig::rms_g4sv21}
\end{figure}

\begin{table}[t]
\caption{\textcolor{blue}{Global RMSE across different mesh numbers and orientation distribution}}
\label{tab:global_RMSE_mesh}
\centering
\resizebox{\columnwidth}{!}{%
\begin{tabular}{l|ccccccccc}
\hline
\textbf{Design/Mesh} & \textbf{2-a} & \textbf{2-r} & \textbf{2-u} & \textbf{8-a} & \textbf{8-r} & \textbf{8-u} & \textbf{18-a} & \textbf{18-r} & \textbf{18-u} \\
\hline
Fig. \ref{fig::pixel_and_s_parameter_first} (a)  & 4.71 & 4.78 & 4.59 & 3.76 & 3.60 & 3.66 & 3.20 & 3.14 & 2.86 \\
Fig. \ref{fig::pixel_and_s_parameter_second} (g) & 3.01 & 2.82 & 2.93 & 2.14 & 2.12 & 2.33 & 2.25 & 2.65 & 2.32 \\
\hline
\end{tabular}%
}
\end{table}

\textcolor{blue}{A table of \ac{RMSE} with respect to different number of triangles per all entry in [S] is shown in Table~\ref{tab:rmse_with_no_of_mesh}, with the diagonal distribution fixed to the alternating case. As mentioned above, \ac{MoM} simulation results with different triangulations per pixel are compared against Ansys HFSS results. In general, the stripline configuration exhibits lower \ac{RMSE} values globally than the microstrip configuration. The microstrip configuration requires more than two triangles per pixel, as the improvement from 2 to 8 triangles exceeds 3~dB, whereas the stripline case improves by only about 0.7–1~dB. Increasing the mesh to 18 triangles is not considered due to the high computational cost of matrix inversion, and the corresponding accuracy gain is negligible. Therefore, 8 triangles per pixel is chosen in the 4-state mirostrip and stripline configuration. It is noted that since 8 triangles per pixel is computationally intractable in 8-state microstrip case, 2 triangles per pixel is chosen.}

\textcolor{blue}{A comparison of the \ac{RMSE} distributions across mesh orientations and densities is shown in Fig.~\ref{fig::rms_IMS_363} and Fig.~\ref{fig::rms_g4sv21} to determine the diagonal distribution. As shown in the previous analysis, 18 triangles per pixel tend to produce almost the same error as compared to the 8-triangle configurations in the case of stripline, while microstrip case shows monotonous decrease but with higher error. This is attributed to the fact that \ac{MoM} implementation is based on the assumption of infinite substrate/ground, while HFSS simulation is simulated with a finite substrate/ground, which means that the solution derived from MoM does not necessarily converge towards the solution derived from HFSS. It is noted that \ac{MoM} shows high discrepancy when it comes to the prediction of the rejection value of $S_{31}$ in the case of Fig.~\ref{fig::rms_IMS_363}.}

\textcolor{blue}{ Clearly, the uniform diagonal pattern is not chosen since the high fluctuation is observed in Fig.~\ref{fig::rms_IMS_363}(d) at $S_{31}$ between 6--7 Ghz, where the other distributions show the error of 2--3~dB. To choose between the alternating and random distribution, Fig~\ref{fig::rms_IMS_363}(e)--(f) and Fig~\ref{fig::rms_g4sv21}(e)--(f) are inspected. Table~\ref{tab:global_RMSE_mesh} summarizes the global \ac{RMSE} values obtained for different mesh densities and orientation distributions. No big global \ac{RMSE} difference is observed between alternating and random cases. However, Fig~\ref{fig::rms_g4sv21}(i) shows an extreme case of stochastic fluctation, evidenced by $S_{32}$ between 7--8 GHz. The error soars up to 8.52 dB. Therefore, it is reasonable to choose 8 traingles per pixel with the alternating distribution for 4-state microstrip/stripline configuration since it shows more stable error.}

In conclusion, the case with eight triangles, demonstrates a balance between accuracy and computational speed. capturing both the notch frequencies and transmission magnitude with reasonable accuracy. \textcolor{blue}{The alternating diagonal distribution was selected because it exhibits stable error behavior without the large fluctuations observed in other configurations. Two triangles per pixel with the alternating distribution was chosen for the 8-state microstrip configuration due to the computationally tractability.}

%% file: sec5.tex
\section{Manufacturing Tolerance}
\label{section::Manufacture_error}   

In this section, study on different configurations including substrate stack-up, the number of trace layers, and lateral and vertical position of the ports are discussed.

\begin{figure}[t!]
    \centering
    \includegraphics[width=1\columnwidth]{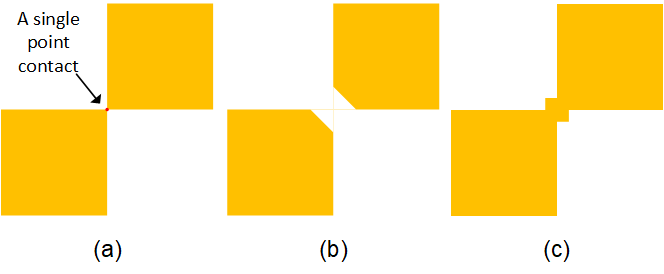}
    \caption{Treatment of two diagonally connected pixels with a single point contact. (a) A single point contact, which is not feasible in a real world. (b) Over-etched two diagonally connected pixels. (c) Under-etched two diagonally connected pixels.}
\label{fig::manufacturing_tolerance}
\end{figure}

\begin{figure}[t!]
    \centering
    \includegraphics[width=1\columnwidth]{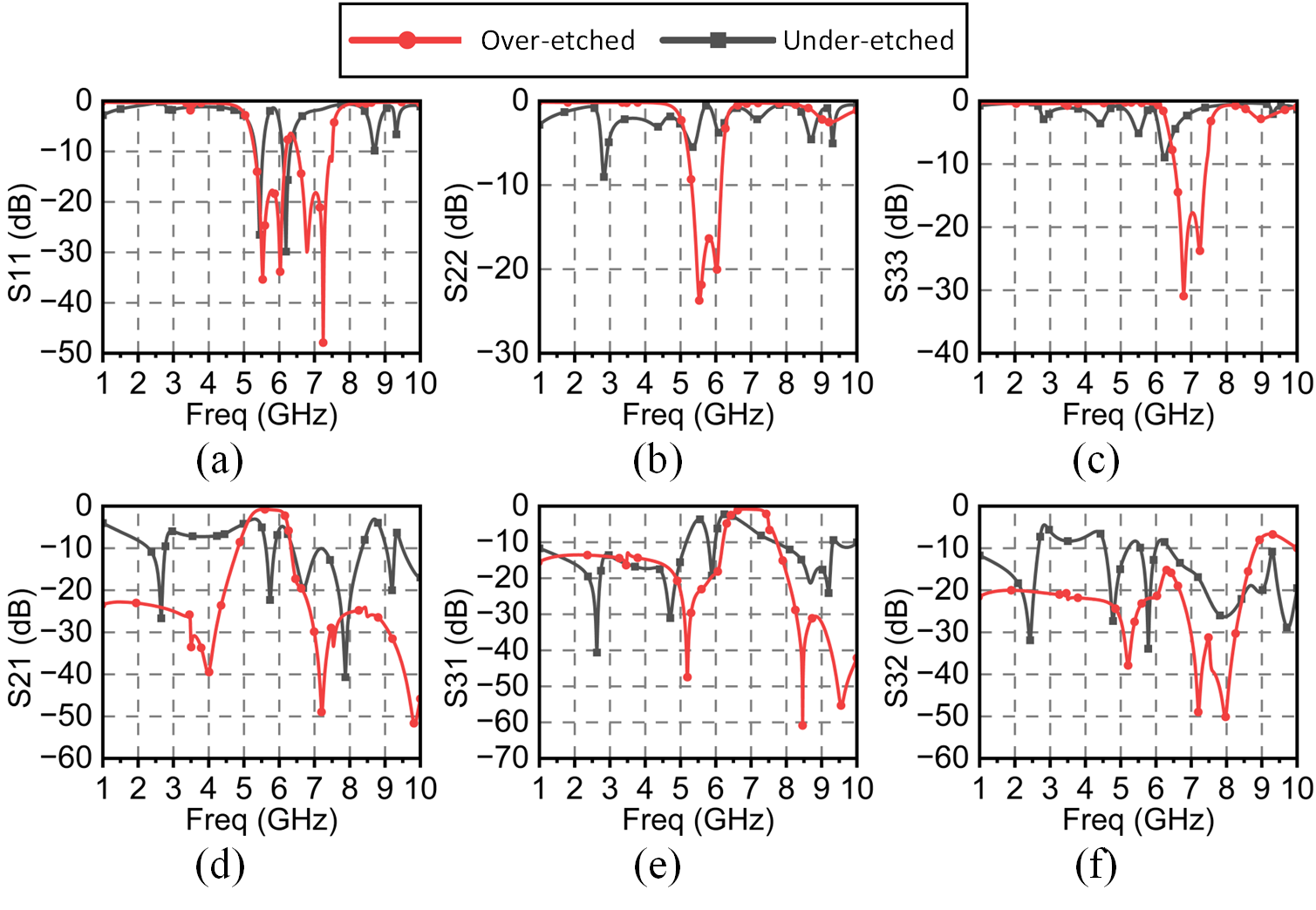}
    \caption{Simulated response of over/under-etched case of Fig. \ref{fig::pixel_and_s_parameter_second}(g).}
\label{fig::tolerance_under_etch_graphs}
\end{figure}

\begin{figure}[t]
    \centering
    \includegraphics[width=1\columnwidth]{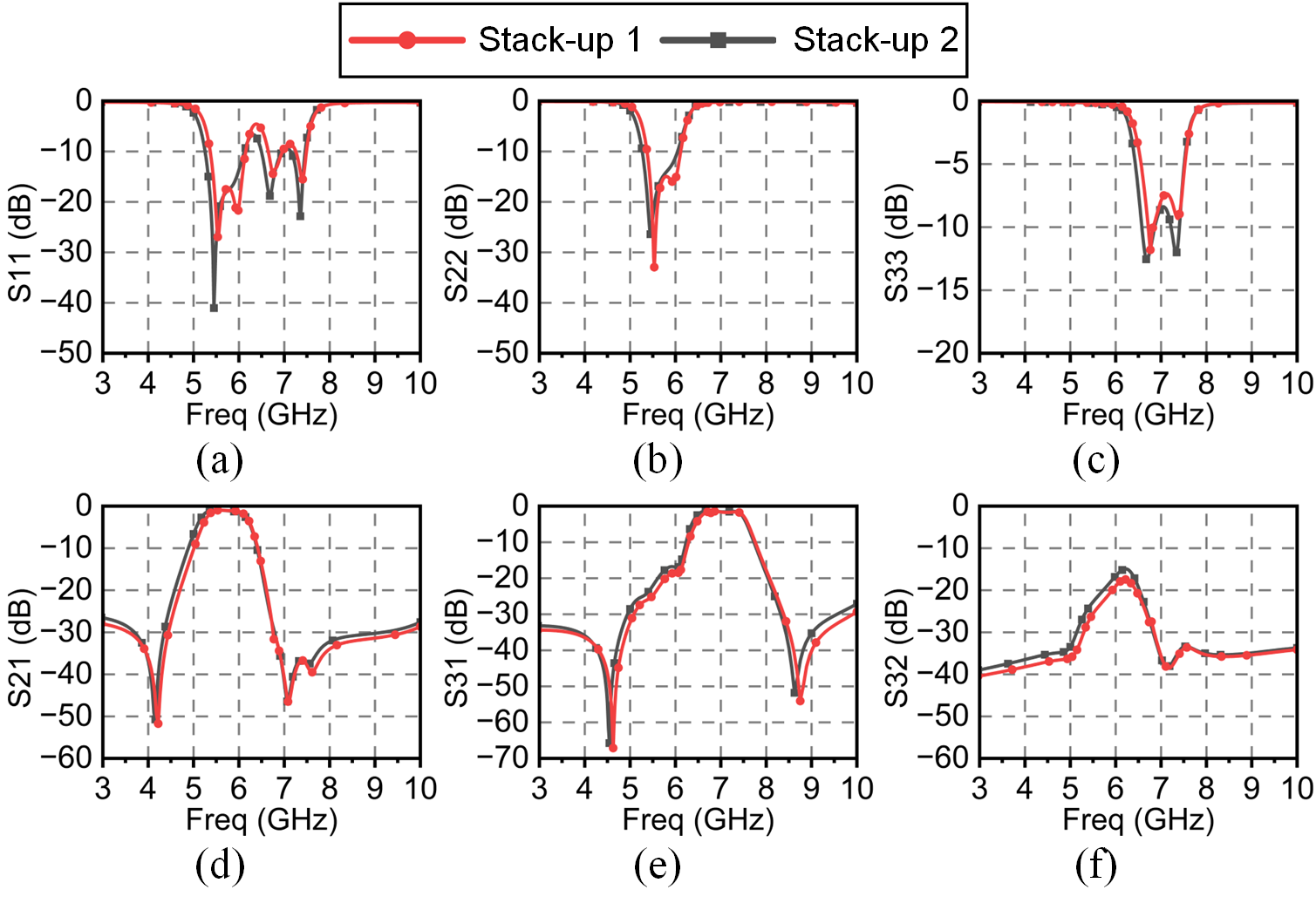}
    \caption{Simulated response of Fig. \ref{fig::pixel_and_s_parameter_second} (h) with different stack-up. Stack-up 1 denotes the original uniform stack-up ($h_1=0.4866mm, h_2=0.4866mm, h_3=0.4866mm$) and Stack-up 2 denotes the non-uniform stack-up ($h_1=0.508mm, h_2=0.426mm, h_3=0.508mm$).}
\label{fig::tolerance_nextpcb_graphs}
\end{figure}

\begin{figure}[t]
    \centering
    \includegraphics[width=1\columnwidth]{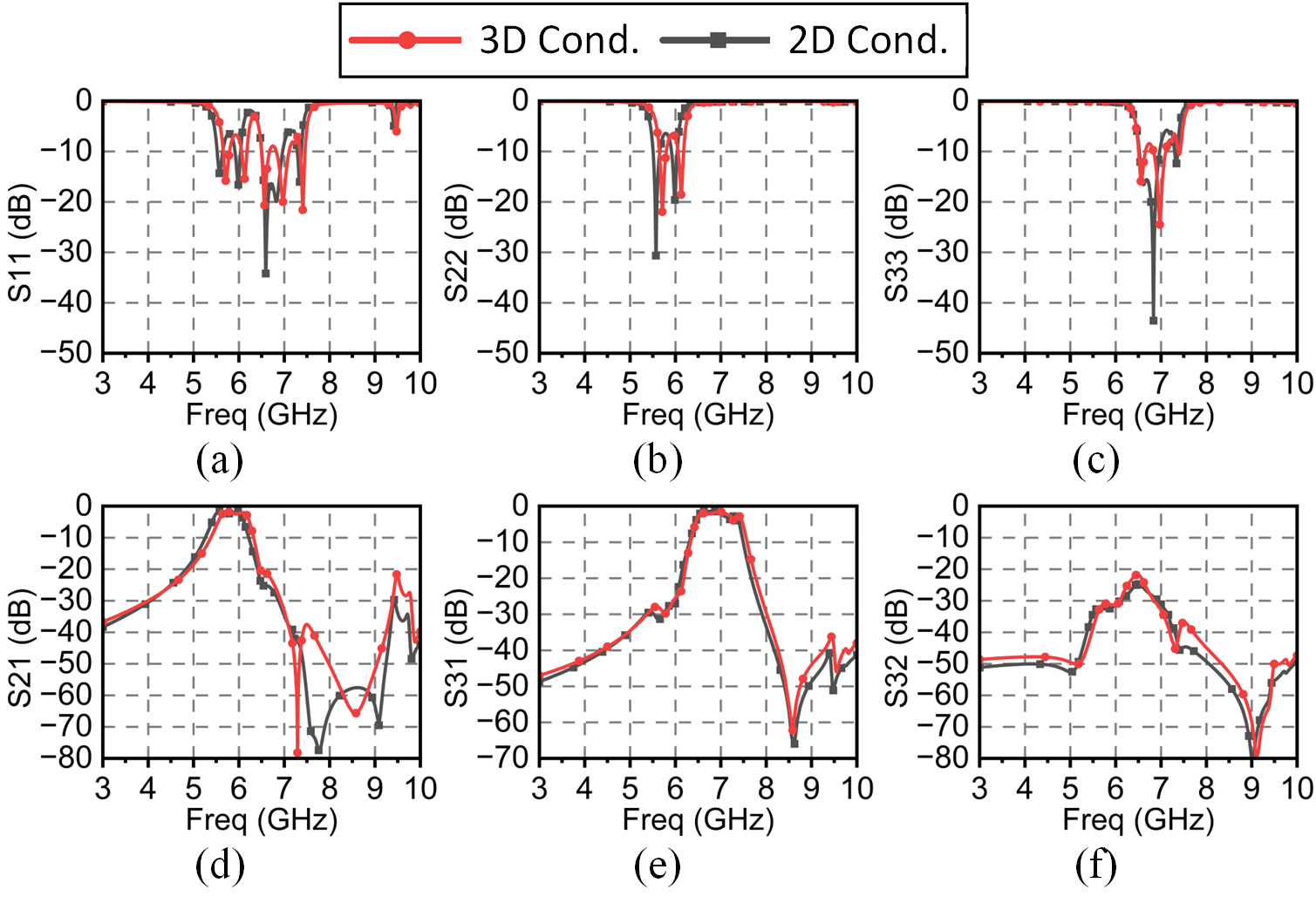}
    \caption{Simulated response of 2D version of Fig. \ref{fig::pixel_and_s_parameter_second}(i).}
\label{fig::tolerance_2d_graphs}
\end{figure}

\subsection{Diagonally Connected Pixels}
\label{subsection::diagonal}

The treatment of diagonally connected pixels, defined as two pixels sharing a single point of contact, requires specific consideration. Possible approaches are illustrated in Fig.~\ref{fig::manufacturing_tolerance}. Since a single-point connection as in Fig.~\ref{fig::manufacturing_tolerance}(a) cannot be fabricated, this work adopts the over-etched configuration shown in Fig.~\ref{fig::manufacturing_tolerance}(b), which aligns with the RWG basis function treatment that considers such pixels electrically isolated. In contrast, most previous studies on pixelated structures either connect the diagonally adjacent pixels as in Fig.~\ref{fig::manufacturing_tolerance}(c) \cite{princetonIMS}, disable those pixels \cite{pixelfilterREF1}, or do not specify their treatment, which further differentiates this work.

\textcolor{blue}{To highlight an importance of taking care of these diagonal pixels, an \ac{RMSE} comparison was performed for the case where a 0.05\,mm square trace is added at the single-point contact, corresponding to Fig.~\ref{fig::manufacturing_tolerance}(c). The under-etched configuration in Fig.~\ref{fig::tolerance_under_etch_graphs} models a fabrication scenario in which pixels intended to remain unconnected become unintentionally connected due to process error. As can be seen in Fig.~\ref{fig::tolerance_under_etch_graphs}, This connection introduces an unintended direct current-flowing path, modifying overall frequency response. Such connections cause severe deviation in the resulting response. In this case, $S_{21}$ shows substantial distortion and loss of band selectivity. It is interesting to note that $S_{31}$ retains partial notch behavior originating from the original geometry. These results indicate that diagonal pixel bridging must be controlled in fabrication.}

\begin{table}[t]
\caption{Substrate Thickness RMSE}
\label{tab:NEXTPCB_rmse}
\centering
\resizebox{\columnwidth}{!}{%
\begin{tabular}{l|cccccc|c}
\hline
\textbf{Sample} & \textbf{S11} & \textbf{S21} & \textbf{S31} & \textbf{S22} & \textbf{S23} & \textbf{S33} & \textbf{Global} \\
\hline
Fig. \ref{fig::pixel_and_s_parameter_second} (g) & 3.83 & 2.92 & 3.89 & 1.41 & 2.95 & 2.52 & 3.04 \\
Fig. \ref{fig::pixel_and_s_parameter_second} (h) & 2.72 & 2.14 & 3.37 & 1.72 & 1.62 & 0.87 & 2.22 \\
Fig. \ref{fig::pixel_and_s_parameter_second} (i) & 2.16 & 6.38 & 2.76 & 1.47 & 3.39 & 1.34 & 3.37 \\
\hline
\end{tabular}%
}
\end{table}

\begin{table}[t]
\caption{2D RMSE}
\label{tab:2d_rmse}
\centering
\resizebox{\columnwidth}{!}{%
\begin{tabular}{l|cccccc|c}
\hline
\textbf{Sample} & \textbf{S11} & \textbf{S21} & \textbf{S31} & \textbf{S22} & \textbf{S23} & \textbf{S33} & \textbf{Global} \\
\hline
Fig. \ref{fig::pixel_and_s_parameter_second} (g) & 4.06 & 2.78 & 2.95 & 1.56 & 2.43 & 2.16 & 2.76 \\
Fig. \ref{fig::pixel_and_s_parameter_second} (h) & 2.57 & 3.03 & 4.17 & 1.66 & 2.25 & 1.01 & 2.64 \\
Fig. \ref{fig::pixel_and_s_parameter_second} (i) & 3.11 & 8.57 & 2.87 & 2.72 & 3.60 & 2.72 & 4.45 \\
\hline
\end{tabular}%
}
\end{table}

\subsection{Substrate Thickness}
\label{subsection::substrate_thickness}

A tolerance study on substrate thickness was performed to evaluate the sensitivity of the optimized design to realistic stack-up variation. Although the design assumes a uniform dielectric thickness, the fabricated board uses Rogers~4350B core and bondply, which introduces slight non-uniformity. Fig.~\ref{fig::tolerance_nextpcb_graphs} compares the response of the original uniform stack-up to a stack-up with small thickness deviations. The two responses remain nearly identical across all S-parameters, indicating that moderate variation in dielectric thickness does not significantly affect device behavior. This result suggests that the diplexer performance is stable under standard PCB manufacturing tolerances with small change in layer thickness.

Table~\ref{tab:NEXTPCB_rmse} presents the \ac{RMSE} analysis for different substrate thickness conditions to evaluate the sensitivity of the diplexer performance to stack-up variations. The overall global RMSE values remain within 2–3.5~dB, indicating that moderate non-uniformity in substrate thickness introduces only minor deviations from the nominal response. A slightly higher RMSE is observed in the $S_{21}$ parameter, particularly in one of the designs, due to its stronger dependence on vertical coupling and layer-to-layer field continuity. Nevertheless, the overall results suggest that the proposed diplexer structures are robust against realistic manufacturing tolerances in dielectric thickness and maintain stable transmission and isolation characteristics across the tested configurations.

\begin{figure}[t]
    \centering
    \includegraphics[width=1\columnwidth]{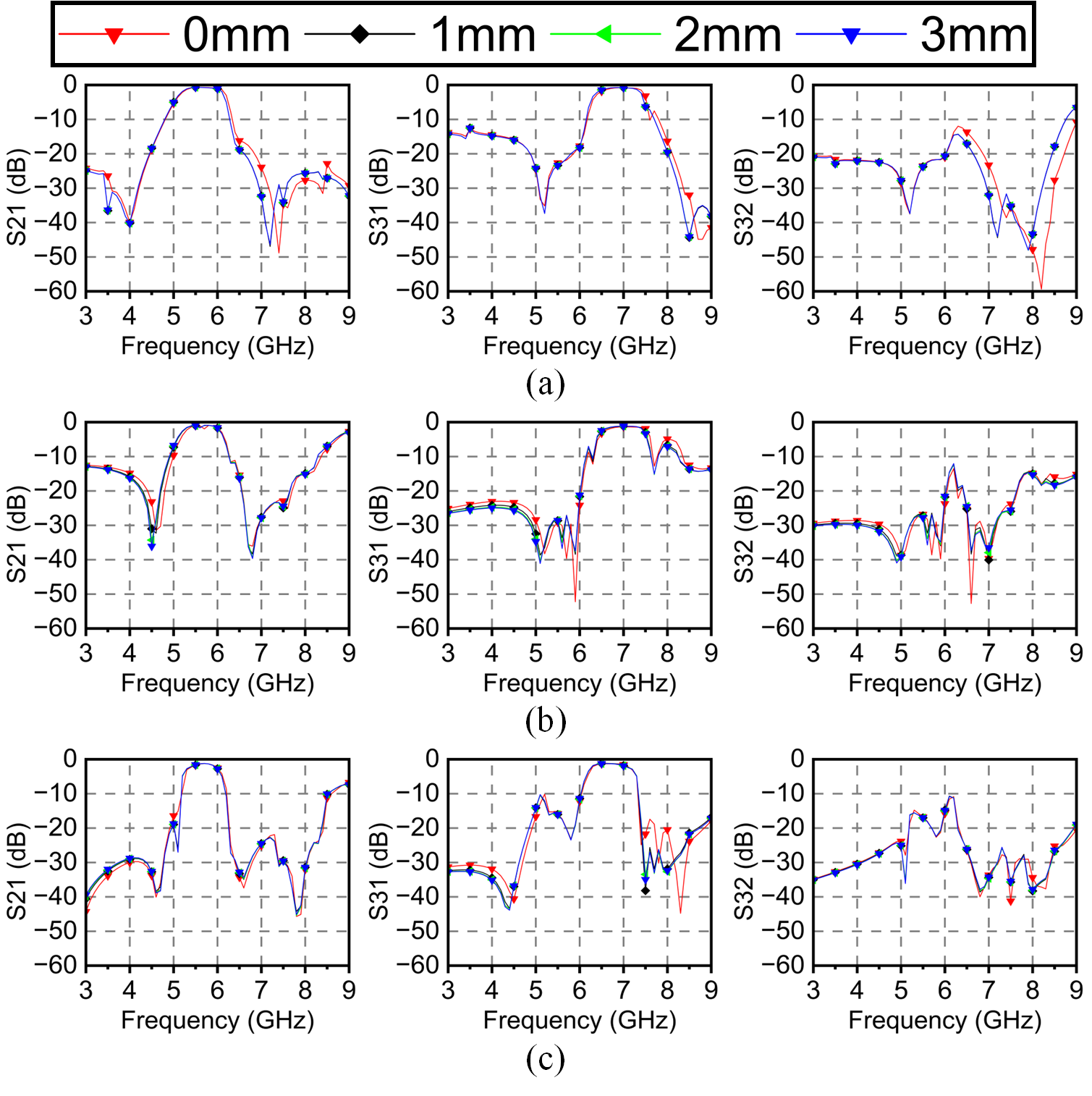}
    \caption{Simulated response of the diplexers with different finite ground size. (a)  Fig. \ref{fig::pixel_and_s_parameter_second}(g). (b) Fig. \ref{fig::pixel_and_s_parameter_first}(g). (c) Fig. \ref{fig::pixel_and_s_parameter_first}(a).}
\label{fig::tolerance_finite_ground_graphs}
\end{figure}

\subsection{2D/3D Conductor}
\label{subsection::2d_3d_metal}

A two-dimensional version of the optimized diplexer layout was simulated to assess the effect of approximating the 3D metal conductors with 2D sheet conductors. As shown in Fig.~\ref{fig::tolerance_2d_graphs}, the 2D model preserves the passband locations, notch positions, and isolation trends observed in the 3D case. This suggests that the dominant coupling mechanism in the examined configurations is lateral surface current interaction, and that vertical conductor thickness and side-wall current paths play a secondary role for these layouts. Minor phase and magnitude deviations appear in certain bands, most notably in the lower band of $S_{21}$ for Fig.~\ref{fig::pixel_and_s_parameter_second}(i), where the absence of side-wall and finite-thickness effects in the 2D approximation becomes more significant.

The \ac{RMSE} values in Table~\ref{tab:2d_rmse} support these observations. For Fig.~\ref{fig::pixel_and_s_parameter_second}(g) and (h), the global \ac{RMSE} remains below 3~dB, indicating that the 2D approximation yields sufficiently accurate results for those geometries. In contrast, the configuration in Fig.~\ref{fig::pixel_and_s_parameter_second}(i) shows a larger error in $S_{21}$, indicating higher dependence on vertical conductor thickness and side-wall current paths. Overall, these results indicate that a 2D surrogate model can serve as a fast \ac{EM} solver for multi-layered pixelated surfaces that have vertical coupling as dominant mechanism, but it may become inadequate, or even impossible, to optimize diplexer configurations where side-wall coupling plays a primary role. This limitation will be addressed in future work.

\subsection{Finite Ground Size}
\label{subsection::finite_ground_size}

Fig.~\ref{fig::tolerance_finite_ground_graphs} evaluates the impact of finite ground-plane extension by increasing $w_{add}$, the ground margin around the pixelated region, from 0\,mm to 3\,mm in Fig.~\ref{fig::tolerance_finite_ground}. The response shows that only the 0\,mm case produces noticeable deviation, while the 1\,mm, 2\,mm, and 3\,mm extensions yield nearly identical characteristics. When the ground plane terminates exactly at the pixel boundary (0\,mm), fringing fields at the aperture edge interact with the surrounding environment, altering the resonant behavior. Once a small ground extension is added, the fringing field is sufficiently confined, and further expansion produces no significant performance change. These results indicate that a small ground margin (approximately 1\,mm) is sufficient to make the frequency response insensitive to surrounding impediments near the diplexers.

\begin{table*}[t]
\centering
\caption{Comparison of the proposed pixelated diplexers with the state-of-the-art diplexers in PCB}
\label{table::comparison}
\setlength{\tabcolsep}{2pt}\renewcommand{\arraystretch}{1.1}
\begin{threeparttable}
\footnotesize
\begin{tabularx}{\textwidth}{@{}C{13mm}|Y|Y|Y|Y|Y|Y|Y|Y|Y|Y|C{17mm}@{}}
\toprule
Ref. & Config.* & Layers & $f_0$ [GHz] & Size ($\lambda_0$) & IL* [dB] & FBW* (\%) & Inter Ch. Rej.* [dB] & Peak Ch. Rej.*  [dB] & Isolation [dB] & Group delay [ns] & Tech. \\
\midrule
\mbox{\cite{TABLE_REF4}} & Microstrip & 1               & 2.45/5.5 & 0.12$\times$0.09  & 1.67/1.58 & 26.9/27.4 & 25/25**  & 32/28**  & 26/30           & N.G  & LC net. \\
\mbox{\cite{TABLE_REF5}} & Microstrip & 2               & 1.8/2.45 & 0.2$\times$0.18   & 2.05/2.15 & 22/40**     & 28/20** & 52/35**  & 30/22**          & N.G  & Slotline \\
\mbox{\cite{TABLE_REF6}}  & SIW        & 1               & 3.5/5.5  & 2.43$\times$1.37  & 3.41/3.4  & 3.4/2.55  & 32/24** & 38/28**  & 33/27.6           & N.G  & Resonator \\
\mbox{\cite{TABLE_REF7}}  & Microstrip & 1               & 1.75/2.35& 0.18$\times$0.34  & 1.34/1.44 & 17/25**     & 22/22** & 40/28**  & 32/25           & N.G  & Resonator \\
\mbox{\cite{TABLE_REF8}}  & Microstrip & 1               & 1.95/2.14& N.G               & 1.12/1.44 & 3/2.8     & 32/34** & 42/58**  & 39/37           & N.G    & Resonator \\
\mbox{\cite{TABLE_REF9}}   & Microstrip & 3               & 1.5/2.75 & 2.39$\times$1.5   & 0.79/1.2  & 62/24     & 32/27** & 55/45**  & 27/28           & 1.79/1.23   & Slotline \\
\mbox{\cite{TABLE_REF10}} & Microstrip & 1               & 2.41/3.61& 1.14$\times$0.26  & 1.46/2.15 & 6.7/3.6   & 37/38** & 45/40**  & 38/38           & N.G    & Resonator \\
\midrule

TW-1 & Microstrip & 3 & 5.7/7 & 0.28$\times$0.2  & 1.2/1.9  & 18/16.7  & 18/23 & 38/56 & 18/16 & 0.91/0.85 & \multirow{2}{*}{%
  \raisebox{-2.0\totalheight}{%
    \parbox[c][0.5\totalheight][c]{17mm}{\centering Multi-layered\\inverse design}%
  }%
} \\
\textcolor{blue}{TW-2} & \textcolor{blue}{Microstrip}  & \textcolor{blue}{3} & \textcolor{blue}{5.9/7} & \textcolor{blue}{0.28$\times$0.2} & \textcolor{blue}{1.65/2.37} & \textcolor{blue}{15/12.4} & \textcolor{blue}{30/29} & \textcolor{blue}{41/45} & \textcolor{blue}{30/25} & \textcolor{blue}{1.1/1.5} & \\
TW-3 & Stripline  & 2 & 5.7/7 & 0.16$\times$0.16 & 0.95/1.44& 19.2/15.8 & 18/19 & 28/44 & 17/20 & 0.86/0.78 & \\
TW-4 & Stripline  & 2 & 5.9/7 & 0.16$\times$0.16 & 1.6/1.5 & 11.3/14.6 & 21/21 & 30/78 & 26/21 & 1.35/1.33 & \\
\bottomrule
\end{tabularx}
\begin{tablenotes}
  \item[*] Config. = configuration, IL = insertion loss, FBW = 3dB fractional bandwidth of IL, Inter Ch. Rej. = Worst inter channel rejection, and Peak Ch. Rej. = Peak inter channel rejection.
  \item[**] Estimated graphically.
  \item[$\dagger$] TW = This work. TW-1 corrresponds to the design in Fig.~\ref{fig::pixel_and_s_parameter_first}(g), TW-2 to the design in Fig.~\ref{fig::FoM_experiment}(b), TW-3 to the design in Fig.~\ref{fig::pixel_and_s_parameter_second}(h), and TW-4 to the design in Fig.~\ref{fig::pixel_and_s_parameter_second}(i). \textcolor{blue}{Fig.~\ref{fig::FoM_experiment}(b) is validated only by the simulation.}
\end{tablenotes}
\end{threeparttable}
\end{table*}

%% file: sec6.tex
\section{Measurement}
\label{sec::measurement}

\begin{figure}[tbp]
    \centering
    \includegraphics[width=1\columnwidth]{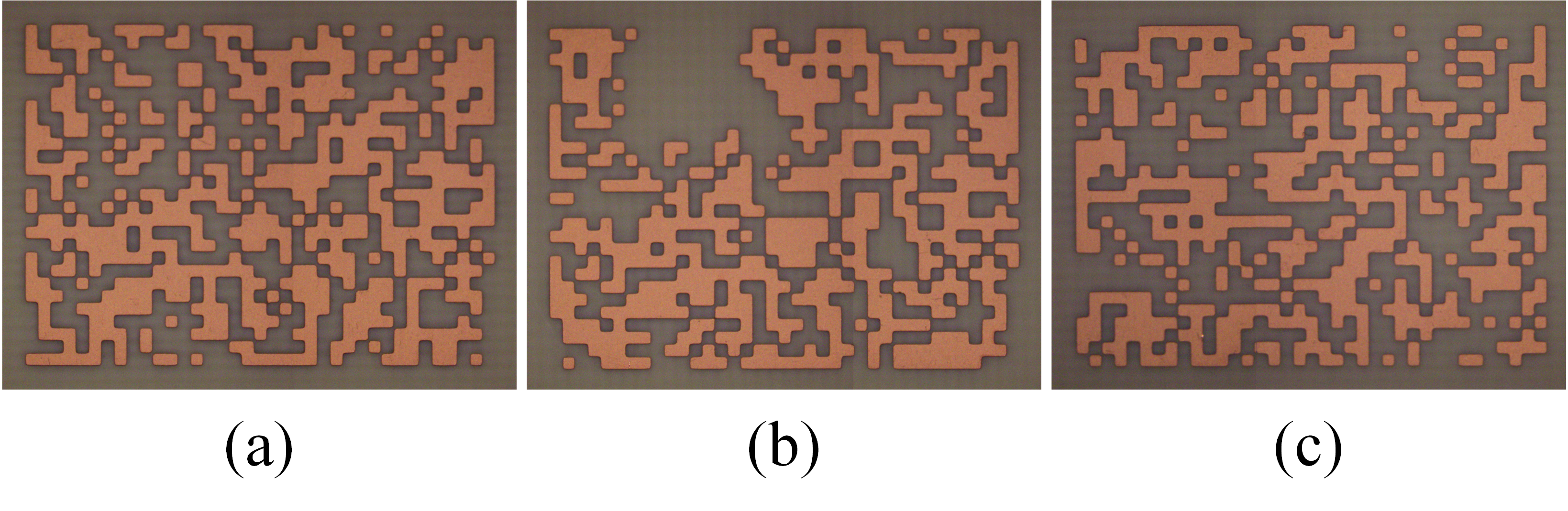}
    \caption{Fabricated diplexers. (a) Fig. \ref{fig::pixel_and_s_parameter_first}(g). (b) Fig. \ref{fig::pixel_and_s_parameter_first}(h). (c) Fig. \ref{fig::pixel_and_s_parameter_first}(i).}
\label{fig::fabrication}
\end{figure}

\begin{figure}[tbp]
    \centering
    \includegraphics[width=1\columnwidth]{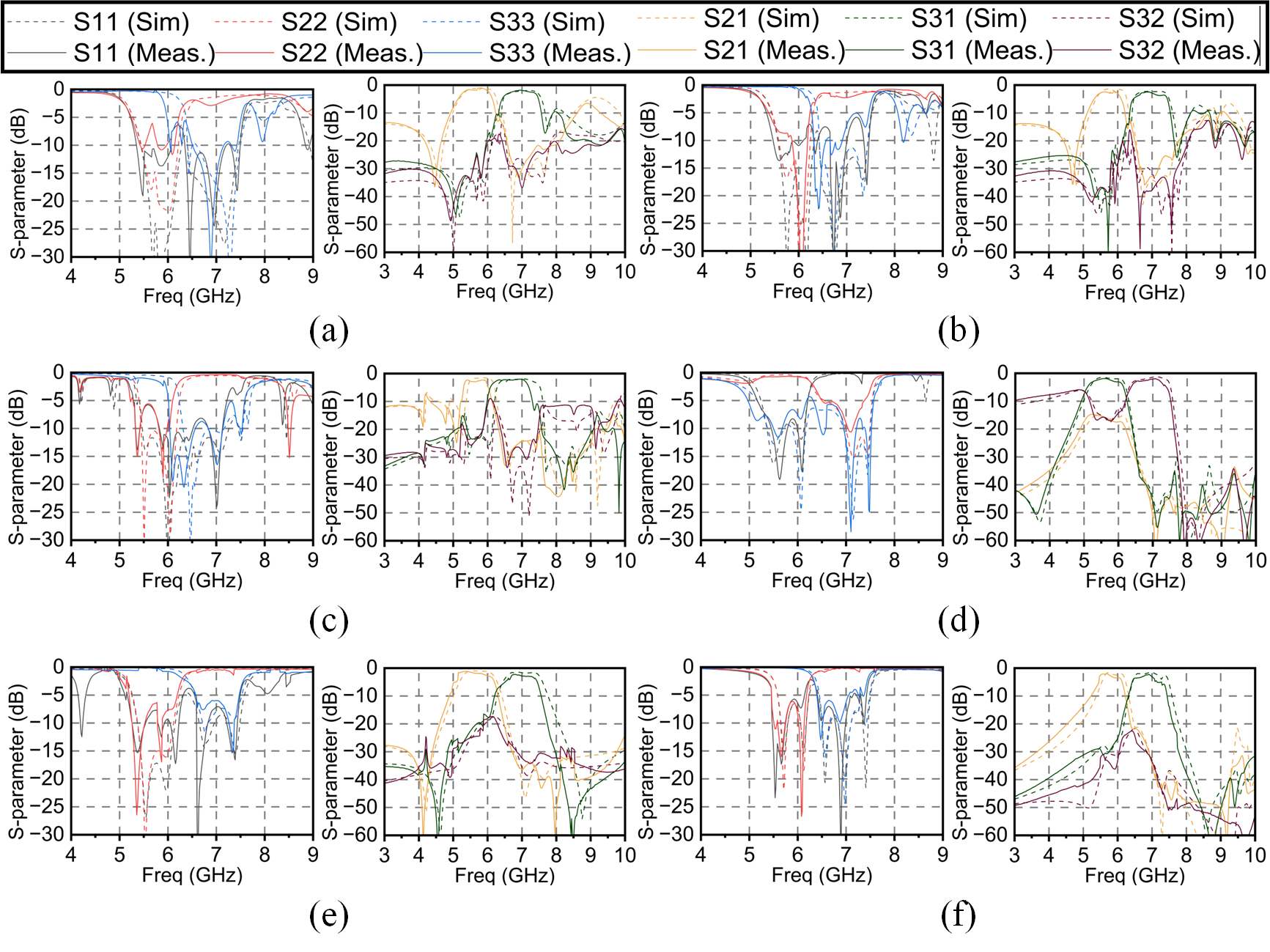}
    \vspace{-6mm}
    \caption{\textcolor{blue}{Measured response. (a)  Fig. \ref{fig::pixel_and_s_parameter_first}(g). (b)  Fig. \ref{fig::pixel_and_s_parameter_first}(h). (c)  Fig. \ref{fig::pixel_and_s_parameter_first}(i). (d) Fig. \ref{fig::pixel_and_s_parameter_second}(c). (e) Fig. \ref{fig::pixel_and_s_parameter_second}(h). (f) Fig. \ref{fig::pixel_and_s_parameter_second}(i).}}
\label{fig::measurement_response}
\end{figure}

\begin{figure}[tbp]
    \centering
    \includegraphics[width=1\columnwidth]{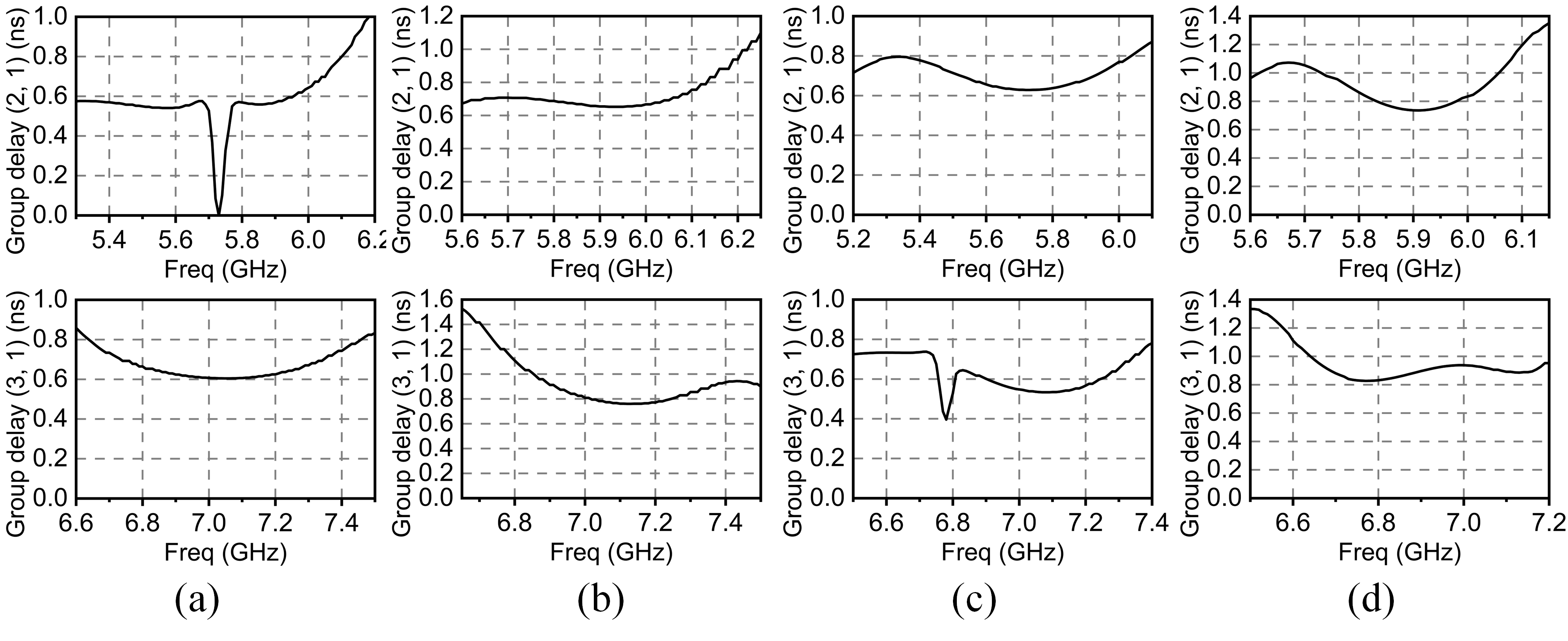}
    \caption{\textcolor{blue}{Simulated group delay response. (a) Fig. \ref{fig::pixel_and_s_parameter_first}(g). (b) Fig. \ref{fig::FoM_experiment}(b). (c) Fig. \ref{fig::pixel_and_s_parameter_second}(h). (d) Fig. \ref{fig::pixel_and_s_parameter_second}(i).}}
\label{fig::group_delay}    
\end{figure}

\renewcommand{\arraystretch}{1.15}
\setlength{\tabcolsep}{3pt}

\renewcommand{\tabularxcolumn}[1]{>{\centering\arraybackslash}m{#1}}
\newcolumntype{Y}{>{\centering\arraybackslash}X}

\renewcommand{\arraystretch}{1.15}
\setlength{\tabcolsep}{3pt}

\begin{table}[t]
\centering
\begin{threeparttable}

\caption{Comparison With Prior Pixelated EM Optimization Works}
\label{tab:comparison_time}

\begin{tabularx}{\linewidth}{c Y Y Y Y Y}
\toprule
\textbf{Ref.} & \textbf{Design} & \textbf{Design Space} & \textbf{Size ($\lambda_0$)} 
& \textbf{EM Sim. + Optim} & \textbf{Total Time} \\
\midrule

\cite{pixelfilterREF1} & Filter 
& $2^{100\times100}$ 
& $0.18\times0.18$ 
& FEM** + GA/BPSO* 
& 19.4\,h--31.4\,h \\

\cite{pixelfilterREF3} & Dual-band filter 
& $2^{32\times32}$ 
& $0.56\times0.56$ 
& FEM** + GAN* 
& N.G + 11 mins \\

{\cite{pixelREF4}} & On-chip PA MN*
& $2^{25\times25}$ 
& $0.07\times0.07$
& MoM** + DNN*
& $>800$ hrs + 8.2 mins \\

\midrule
This work & Diplexer 
& \begin{tabular}{@{}c@{}}
$8^{27\times37}$\\[2pt]
$4^{30\times30}$
\end{tabular}
& \begin{tabular}{@{}c@{}}
$0.28\times0.2$\\[2pt]
$0.16\times0.16$
\end{tabular}
& Pre-computed MoM + Tree-Search 
& 2 hours + 45 mins; 4 hours + 6 hours 51 mins \\
\bottomrule
\end{tabularx}

\begin{tablenotes}[flushleft]
\footnotesize
\item[*] GA = genetic algorithm, BPSO = binary particle swarm optimization, GAN = generative adversarial network, DNN = deep neural network, and MN = matching network
\item[**] FEM = Ansys HFSS in \cite{pixelfilterREF1} and \cite{pixelfilterREF3}, MoM = Advanced Design System Momemtum in \cite{pixelREF4}
\end{tablenotes}
\end{threeparttable}
\end{table}

\textcolor{blue}{The fabricated diplexers, shown in Fig.~\ref{fig::fabrication}, were manufactured using a standard PCB process with a minimum trace gap of 0.1~mm to minimize the number of any unintended under-etching. Despite this constraint, some diagonally connected pixels are still observed in Fig.~\ref{fig::fabrication}, which introduces slight deviations in the measured results. The measurements were carried out using a Rohde \& Schwarz ZVA67 VNA, and connector de-embedding was performed following the procedure in \cite{deembedding}. The selected measurement results are shown in Fig.~\ref{fig::measurement_response}. Although the designs were optimized assuming a uniform substrate stack-up, the fabricated structures use Rogers 4350B with bondply, resulting in a slightly non-uniform dielectric profile, as described in the caption of Fig.~\ref{fig::tolerance_nextpcb_graphs}. Nevertheless, the measured results show strong agreement with simulations in terms of passband bandwidth and notch frequencies, indicating good tolerance to practical fabrication variations. The stripline configuration exhibits better tolerance in passband bandwidth and frequency of maximum suppression than the microstrip case, which is attributed to the use of fewer conductive traces due to the reduced layer complexity and and more symmetric vertical couplings.}

The simulated group delay responses of the optimized diplexers are shown in Fig.~\ref{fig::group_delay}. While designs of Fig.~\ref{fig::FoM_experiment}(b) and Fig.~\ref{fig::pixel_and_s_parameter_second}(i) exhibit a relatively flat group delay across the passband, others show pronounced non-uniformity, including a localized sharp irregularity consistent with a parasitic resonance. This behavior is primarily attributed to parasitic resonances within the passband.

Table~\ref{table::comparison} provides a comprehensive comparison between the proposed pixelated diplexers and previously reported PCB-based diplexers and pixelated filters. The worst attenuation from the input port to the other-channel port at the first band, followed by that of the subsequent band, is reported. Isolation is defined as the worst-case attenuation between ports other than the common input port in two passbands. Group delays in the table denotes the maximum group delay in the first and second bands within 1~dB bandwidth. The most difficult specifications to be satisfied was inter-channel rejections below -20 dB throughout the passbands, which was achieved in the diplexers in Fig.~\ref{fig::pixel_and_s_parameter_second}(i) while trading off bandwidth, group delays and insertion loss. It is highlighted that this work aims to achieve the dividing and filtering response in the single foot-print, which was rarely done in the conventional design process due to the intractability of the design. Additional performance gains—particularly in isolation and inter-channel rejection—are anticipated with the introduction of via walls as additional field-confinement structures, the use of shorting vias inside the pixel map, and cascade of the band-pass filters and dividers.

Table~\ref{tab:comparison_time} summarizes a comparison between the proposed optimization framework and prior pixelated EM design approaches. Compared with conventional full-wave–driven global optimization methods, the proposed pre-computed MoM–based tree-search framework achieves faster optimization time while exploring high-dimensional design spaces. In contrast to neural-network-driven techniques, although neural models provide faster inference time or \ac{EM} simulation time, they introduce considerable overhead associated with data acquisition and training.


%% file: conclusion.tex
\section{Conclusion}
\label{sec:conclusion}

This paper introduced a computationally efficient inverse design framework that extends pixelated EM synthesis to multi-layered, multi-port diplexers for the first time, built upon pre-computed, rigorous electromagnetic analysis of \ac{MoM}. By integrating a pre-computation-based MoM solver with a depth-increasing all-way tree search algorithm, the method achieves rapid optimization across large design spaces while maintaining reasonable accuracy. The resulting diplexers, implemented in both microstrip and stripline forms, demonstrate low insertion loss, moderate inter-channel rejection/isolation, and compact geometry in \ac{UNII} bands, confirming the feasibility of scalable white-box inverse design. The framework’s robustness was further verified through systematic studies of meshing, and tolerance study on a stack-up, finite ground size, and under-etch effects, revealing that the designs remain stable against realistic PCB manufacturing tolerances. Future works include incorporating lumped components, additional states with vias, and diplexers cascaded with multiple inversely designed passive surfaces.